\documentclass[preprint,10pt]{elsarticle}
\usepackage{lineno,hyperref}
\usepackage{graphicx}
\usepackage{caption2}
\usepackage{amsmath}
\usepackage{amssymb}
\usepackage{bbding}
\usepackage{graphics}
\usepackage{pifont}
\usepackage{bm}
\usepackage{amsfonts}
\usepackage{mathrsfs}
\usepackage{float}
\usepackage{epstopdf}
\usepackage[footnotesep=0.4in]{geometry}
\usepackage{subfigure}
\biboptions{sort&compress}
\modulolinenumbers[5]

\makeatletter
\newcommand\figcaption{\def\@captype{figure}\caption}
\newcommand\tabcaption{\def\@captype{table}\caption}
\DeclareMathOperator{\sech}{sech}

\DeclareMathOperator{\e}{e}
\newcommand{\PT}{{\cal PT}}
\def\d{\,\mathrm{d}}
\allowdisplaybreaks

\begin{document}

\begin{frontmatter}

\title{Stability analysis of nonlinear localized modes in the coupled Gross-Pitaevskii equations with $\PT$-symmetric Scarf-II potential}

\author[mainaddress]{Jia-Rui Zhang\corref{correspondingauthor}}
\cortext[correspondingauthor]{Corresponding author}
\ead{jrzhang@cau.edu.cn}

\author[mainaddress]{Xia Wang}

\address[mainaddress]{College of Science, China Agricultural University, Beijing 100083, China}

\begin{abstract}
We study the nonlinear localized modes in two-component Bose-Einstein condensates with parity-time-symmetric Scarf-II potential, which can be described by the coupled Gross-Pitaevskii equations. Firstly, we investigate the linear stability of the nonlinear modes in the focusing and defocusing cases, and get the stable and unstable domains of nonlinear localized modes. Then we validate the results by evolving them with $5\%$ perturbations as an initial condition. Finally, we get stable solitons by considering excitations of the soliton via adiabatical change of system parameters. These findings of nonlinear modes can be potentially applied to physical experiments of matter waves in Bose-Einstein condensates.
\end{abstract}


\end{frontmatter}


\section{Introduction}
\label{sec-1}

Bose-Einstein condensate (BEC)~\cite{Einstein1924,Einstein1925}, as one of the important physical phenomena, has attracted the attention of researchers. The successful observation of solitons in BECs has become one of the research focuses in the fields of condensed matter physics and atom optics~\cite{Anderson1995,Davis1995,Strecker2002}. Compared with the single-component ones, the multi-component BECs possess the inter-component interactions and have complicated quantum phases and properties~\cite{Myatt1997,Pu1998,Modugno2001,Alon2005,Kapale2005,Bhongale2008,Zawadzki2010,Zoest2010}. Many novel phenomenons have been discovered in multi-component BECs~\cite{Victor2003,Gaspar2004,Kasamatsu2004,Malomed2004,Brazhnyi2005,Kevrekidis2005,Victor2005,Adhikari2005,Xue2008}, including symbiotic solitons, soliton trains, soliton pairs, multi-domain walls, and multi-mode collective excitations. As one kind of multi-component BECs, the two-component BECs trapped in a quasi-one-dimensional harmonic potential at zero temperature can be described by the following coupled Gross-Pitaevskii equations~\cite{Ho1996,Esry1997}:
\begin{subequations}\label{eq1}
\begin{align}
i\hbar\frac{\partial \Psi_1}{\partial t}&=\Big(-\frac{\hbar^2}{2 M}\nabla^2+V_1(\bm{r})+g_{11}|\Psi_1|^2+g_{12}|\Psi_2|^2) \Big) \Psi_1,\ \\
i\hbar\frac{\partial \Psi_2}{\partial t}&=\Big(-\frac{\hbar^2}{2 M}\nabla^2+V_2(\bm{r})+g_{21}|\Psi_1|^2+g_{22}|\Psi_2|^2) \Big) \Psi_2,\
\end{align}
\end{subequations}
where $\Psi_j$ is the two-component parameter, $\bm{r} = (x,y,z)$, $\hbar$ is the Planck constant, $M$ is the atomic mass, $\nabla^2$ is the Laplacian, $V_j(\bm{r})=[\omega_{jx}^2x^2/2+\omega_{j\bot}^2(y^2+z^2)]M/2$ stands for the harmonic potentials, $g_{jj}$ is the interactions between atoms, and $g_{j,3-j}$ describes the inter-component interactions ($j=1,2$)~\cite{Liu2009,Zhang2009,Yakimenko2012,Wang2013}.

If the trap frequencies in the radial directions $\omega_{j\bot}$ are larger than the axial directions $\omega_{jx}$, Eq.~(\ref{eq1}) becomes a quasi-one-dimensional system along the $x$ direction. Through the normalization and transformation $\Psi_j\rightarrow\psi_j$, $x\rightarrow\sqrt{\hbar/(M\omega_{1\bot})}x$, $t\rightarrow(2\pi/\omega_{1\bot})t$, Eq.~(\ref{eq1}) can be written in the form~\cite{Liu2009,Zhang2009,Yakimenko2012,Wang2013}:
\begin{subequations}
\begin{align}
i\frac{\partial \psi_1}{\partial t}&=\Big(-\frac{1}{2}\frac{\partial^2}{\partial x^2}+\frac{\lambda_1^2}{2}x^2+b_{11}|\psi_1|^2+b_{12}|\psi_2|^2) \Big) \psi_1,\ \\
i\frac{\partial \psi_2}{\partial t}&=\Big(-\frac{1}{2}\frac{\partial^2}{\partial x^2}+\frac{\lambda_2^2}{2}x^2+b_{21}|\psi_1|^2+b_{22}|\psi_2|^2) \Big) \psi_2,\
\end{align}
\end{subequations}
where $\lambda_j=\omega_{jx}/\omega_{j\bot}$, $b_{jj}$ and $b_{j,3-j}$ are related to $\int|\psi_j|^2\d x$, trap frequencies in the radial directions $\omega_{j\bot}$, and interactions between atoms $g_{jj}$ or inter-component $g_{j,3-j}$ ($j=1,2$).
For the harmonic potentials, the soliton states were studied widely~\cite{Liu2009,Zhang2009,Yakimenko2012,Wang2013}. In this paper, we consider the two-component BECs trapped in another potential.

Put forward by Bender and his coworker in 1998~\cite{Bender1998,Bender2003,Bender2007}, parity-time- ($\PT$-) symmetry behaviors have attracted much attention in both non-Hermitian Hamiltonian systems and nonlinear wave systems~\cite{Nixon2012,Lumer2013,Yang2014}, which make systems with complex potentials possibly support fully-real linear spectra~\cite{Ahmed2001} and stable nonlinear modes~\cite{Makris2008,Yan2015,yan2015,Chen2018}. That is, the potential function $U(x) = V(x) + iW(x)$ satisfies $V(x) = V(-x)$ and $W(-x) = -W(x)$~\cite{Bender1998,Bender2003,Bender2007}. Over the past few years, various $\PT$-symmetric potentials have been introduced into the nonlinear Schr\"{o}dinger equation and the existence of different nonlinear local modes is analytically and numerically investigated~\cite{Shi2011,Hu2012,Khare2012,Achilleos2012,Yan2013,Shen2018,Xu2020,Song2022,Zhong2023}. To better investigate physical phenomena, it is meaningful to introduce new forms of $\PT$-symmetric potentials in nonlinear systems.

In this paper, we investigate the coupled Gross-Pitaevskii equations with complex $\PT$-symmetric potentials~\cite{Wang2013}:
\begin{subequations}\label{equation}
\begin{align}
i\frac{\partial \psi_1}{\partial t}&=\Big(-\frac{\partial^2}{\partial x^2}-a_1(|\psi_1|^2+|\psi_2|^2) - U_1(x) \Big) \psi_1,\ \\
i\frac{\partial \psi_2}{\partial t}&=\Big(-\frac{\partial^2}{\partial x^2}-a_2(|\psi_1|^2+|\psi_2|^2) - U_2(x) \Big) \psi_2,\
\end{align}
\end{subequations}
where $a_j$ represent the assumed-equal intra-component and inter-component interactions, $U_j(x)$ are the complex $\PT$-symmetric potentials, and the imaginary parts of $U_j(x)$ stand for the gain or loss term from the thermal clouds.

The present paper is built up as follows. In Sect.~\ref{sec-2}, we consider the analytic bright-soliton solution in the coupled Gross-Pitaevskii equations with complex $\PT$-symmetric Scarf-II potentials; the linear stability analysis and the numerical evolution results corroborating the analytical solitons are presented in Sect.~\ref{sec-3}; In Sect.~\ref{sec-4}, we perform numerical simulations for the excitation and evolution of nonlinear modes via adiabatical change of system parameters; Finally, conclusions and discussions are given in Sect.~\ref{sec-5}.

\section{Localized modes in coupled Gross-Pitaevskii equations}
\label{sec-2}

We concentrate on stationary solutions of Eq.~(\ref{equation}) in the form
\begin{equation}
\psi_j(x,t) =\phi_j(x) \e^{i \nu_j t},\quad j=1,2 \ ,
\end{equation}
where $\nu_j$ are the real propagation constant. The complex solutions $\phi_j(x)$ satisfy the following condition
\begin{subequations}\label{4thcqNLS}
\begin{align}
\big[ \frac{d^2}{d x^2} + a_1 (|\phi_1|^2+|\phi_2|^2) + U_1(x) \big] \phi_1 = \nu_1 \phi_1\ ,\\
\big[ \frac{d^2}{d x^2} + a_2 (|\phi_1|^2+|\phi_2|^2) + U_2(x) \big] \phi_2 = \nu_2 \phi_2\ ,
\end{align}
\end{subequations}
which can be solved for the given potentials $U_j(x)$ and real propagation constant $\nu_j$.

For the $\PT$-symmetric potentials $U_j(x)$ are all chosen as the well-known Scarf-II potentials as
\begin{equation}
U_j(x)=V_j\sech^2(x)+i W_j\sech(x)\tanh(x).
\end{equation}

We have the analytic bright-soliton solution as~\cite{Yan2015}
\begin{equation}\label{jie}
\phi_j(x) = A_j\, \text{sech} (x) \e^{i \varphi_j},\quad j=1,2 \ ,
\end{equation}
with the phase is
\begin{equation}
\varphi_j = \frac{W_j}{3} \arctan[\sinh(x)],\quad j=1,2 \ ,
\end{equation}
under the constraints of
\begin{equation}\label{jie1}
\frac{18+W_j^2-9 V_j}{9 a_j} = \sum\limits_{n=1,2} A_n^2,\quad j=1,2 \ ,
\end{equation}
and $\nu_j = 1$.

For the nonlinear modes given in Eq.~(\ref{jie}), the power of the solutions is defined as $P_j=\int_{-\infty}^{\infty}|\phi_j(x)|^2dx$, $P=P_1+P_2$, while the Poynting vector $S_j = \frac{i}{2}(\phi_j\phi_{jx}^* - \phi_j^*\phi_{jx})=A_j^2W_j/3\, \text{sech}^3(x)$. The power flows from left (right) to right (left) at $x_0$ when $S(x_0)>0$ ($S(x_0)<0$).

\section{Linear Stability analysis}
\label{sec-3}

In this section, we investigate the linear stability of the nonlinear modes, which is a standard protocol to show the stability of nonlinear localized modes. We consider the perturbed solution $\psi_j(x,t)$, in the form
\begin{equation}\label{pertub}
\psi_j(x,t) = \phi_j(x) \e^{i \nu_j t} +\epsilon \, \big[f_j(x) \e^{i \delta t} +g_j^*(x) \e^{-i \delta^* t} \big] \e^{i \nu_j t} \,
\end{equation}
where $\epsilon \ll 1$, which is the small perturbation on the solution. $f_j(x)$ and $g_j(x)$ are the perturbation eigenfunctions of the linearized eigenvalue problem. By substituting Eq.~(\ref{pertub}) into Eq.~(\ref{4thcqNLS}) and linearizing with respect to $\epsilon$, we can drive the following linear eigenvalue problem:
\begin{eqnarray}\label{linstab}
\left(
  \begin{array}{cccccc}
    L_1 & a_1\phi_1^2 & a_1 \phi_1 \phi_2^* & a_1 \phi_1 \phi_2 \vspace{0.05in}\\
    -a_1 \phi_1^{*2} & -L_1^* & -a_1 \phi_1^* \phi_2^* & -a_1 \phi_1^* \phi_2 \vspace{0.05in}\\
    a_2 \phi_1^* \phi_2 & a_2 \phi_1 \phi_2 & L_2 & a_2 \phi_2^2 \vspace{0.05in}\\
    -a_2 \phi_1^* \phi_2^* & -a_2 \phi_1 \phi_2^* & -a_2 \phi_2^{*2} & -L_2^*
  \end{array}
\right) \left(
         \begin{array}{c}
           f_1 \vspace{0.05in}\\
           g_1 \vspace{0.05in}\\
           f_2 \vspace{0.05in}\\
           g_2
         \end{array}
       \right)=\delta \left(
         \begin{array}{c}
           f_1 \vspace{0.05in}\\
           g_1 \vspace{0.05in}\\
           f_2 \vspace{0.05in}\\
           g_2
         \end{array}
       \right),
\end{eqnarray}

where
\begin{subequations}
\begin{align}
L_1 = -\nu_1 + \partial_x^2 +U_1 + 2 a_1 \phi_1 \phi_1^* + a_1 \phi_2 \phi_2^* \, \\
L_2 = -\nu_2 + \partial_x^2 +U_2 +   a_2 \phi_1 \phi_1^* + 2 a_2 \phi_2 \phi_2^* \,
\end{align}
\end{subequations}

The imaginary part of $\delta$ measures the growth rate of the perturbation instability. If $|\text{Im}(\delta)| > 0$, then the perturbation will grow exponentially with $t$, and the solutions are unstable; otherwise, the solutions are stable. In our numerical simulation, we use the Fourier collocation method to discretize the associated differential operator as a matrix to solve the eigenvalue problem~\cite{Trefethen2000}. To further verify the stability of the solitons, we numerically investigate the stability by evolving them with $5\%$ perturbations as the initial condition to simulate the random white noise (i.e., $\psi(x,0)=\phi(x)(1+\xi)$ and $\xi$ represents $5\%$ perturbations). In our numerical simulations, the second-order spatial differential is carried out by using Fourier spectral collocation method, and the integration in time is carried out by using the explicit fourth-order Runge-Kutta method~\cite{yang2010}.

\begin{figure}[htbp]
\begin{center}
\renewcommand{\captionfont}{\scriptsize}
\renewcommand{\captionlabelfont}{\scriptsize}
\renewcommand{\figurename}{Figs.\,}
\subfigure[]{
\includegraphics[scale=0.57]{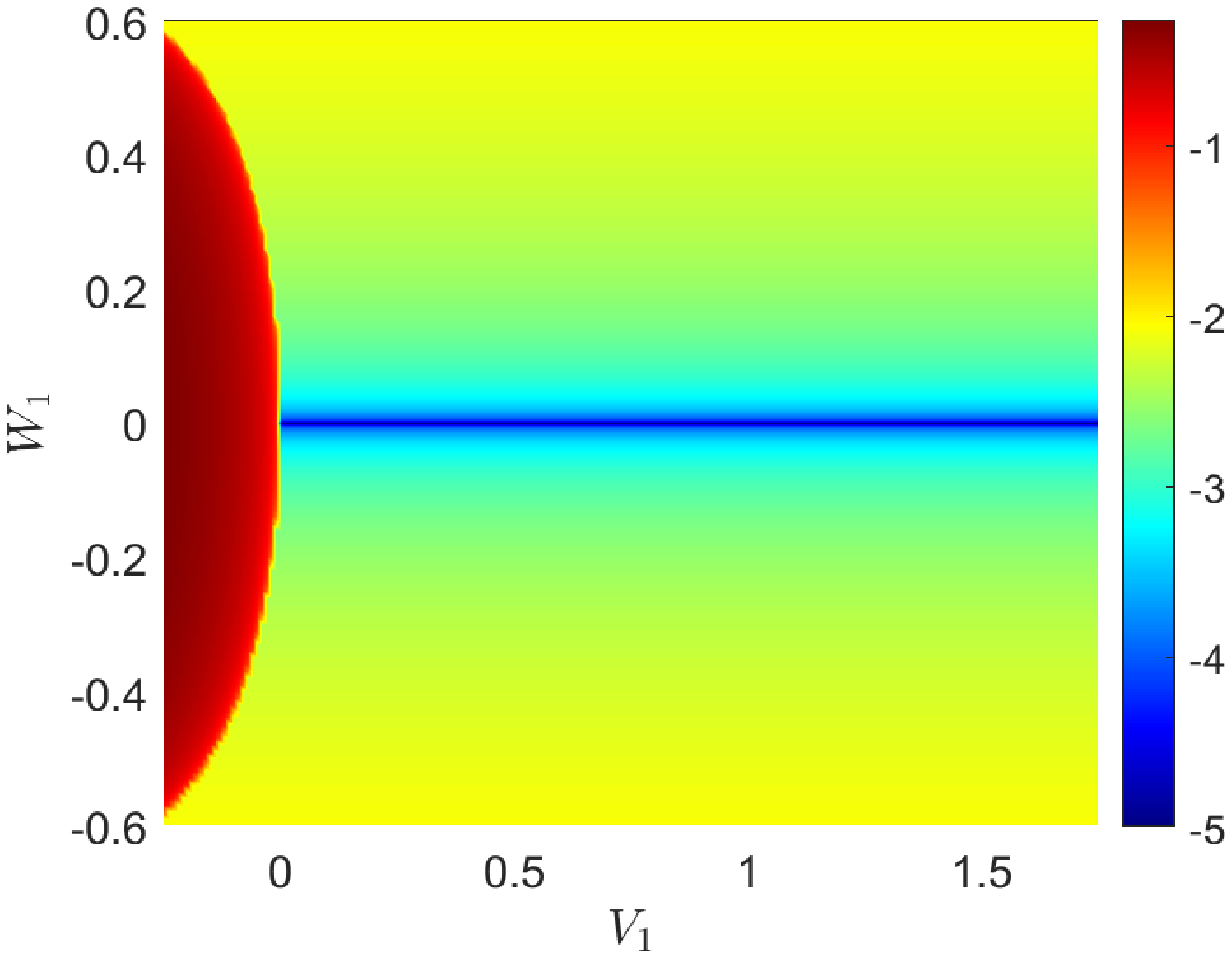}}
\subfigure[]{
\includegraphics[scale=0.57]{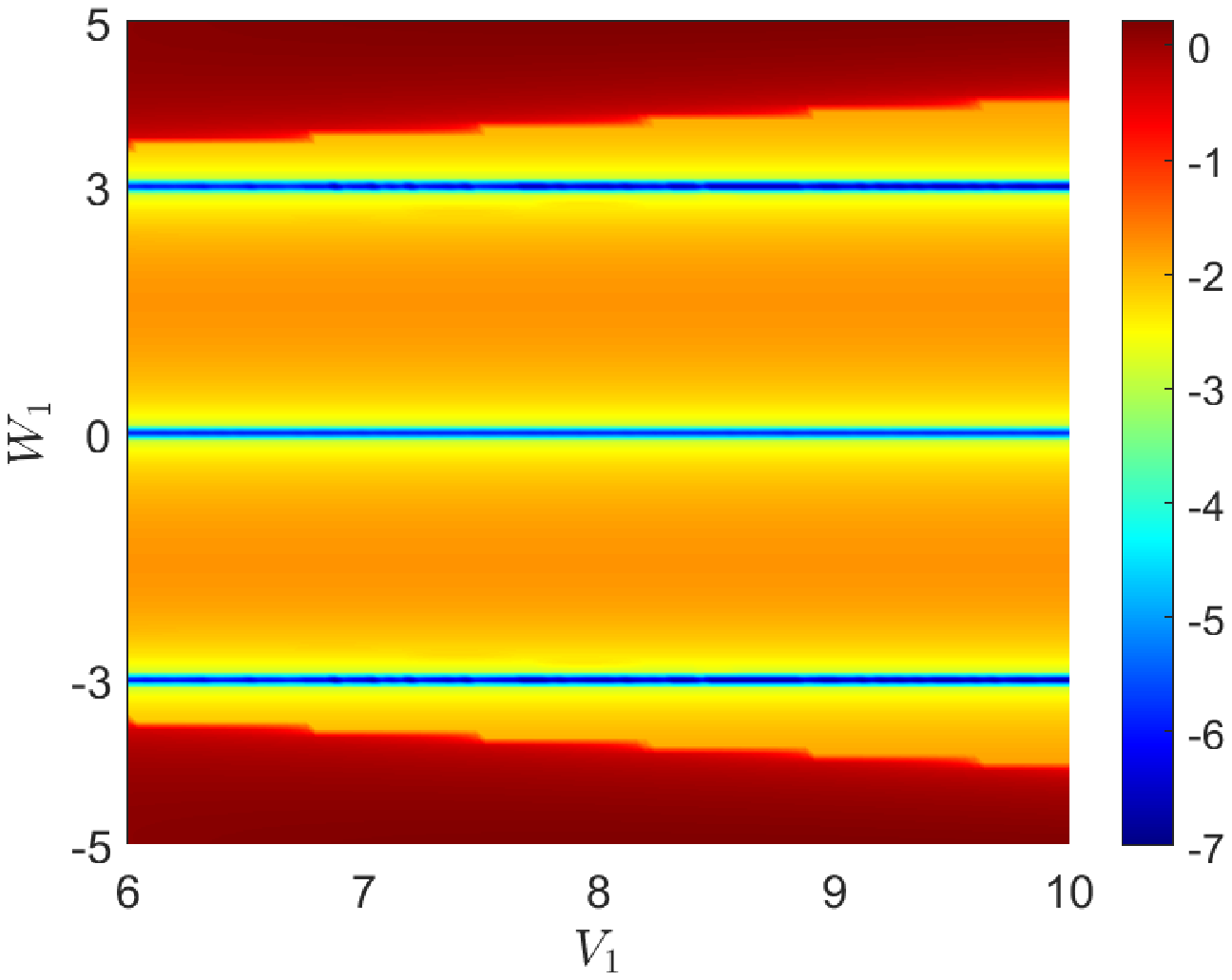}}
\end{center}
\figcaption{\footnotesize
Maximal imaginary part of the linearization eigenvalue $\delta$ in the ($V_1$, $W_1$)-space (common logarithmic scale), under the constraint of $A_1 = 0.5$, $V_1 = V_2$, $W_1 = W_2$ and (a) $a_1 = 1$; (b) $a_1 = -1$.
}
\label{1}
\end{figure}

Firstly, under the constraint of $A_1 = 0.5$, $V_1 = V_2$, $W_1 = W_2$, we consider the focusing case $a_1 = a_2 = 1$ and the defocusing case $a_1 = a_2 = -1$, respectively. Then we get the stable (blue) and unstable (red) domains of nonlinear localized modes in $(V_1, W_1)$ space [see Figs.~\ref{1}]. They are determined by the maximum absolute value of imaginary parts of the linearized eigenvalue $\delta$ in Eq.~(\ref{linstab}). We find that solitons tend to be unstable with the increase of $|W_1|$ in the focusing case. It is worth noting that when $|W_1|=3$, solitons are stable in the defocusing case.

\begin{figure}[htbp]
\begin{center}
\renewcommand{\captionfont}{\scriptsize}
\renewcommand{\captionlabelfont}{\scriptsize}
\renewcommand{\figurename}{Figs.\,}
\subfigure[]{
\includegraphics[scale=0.57]{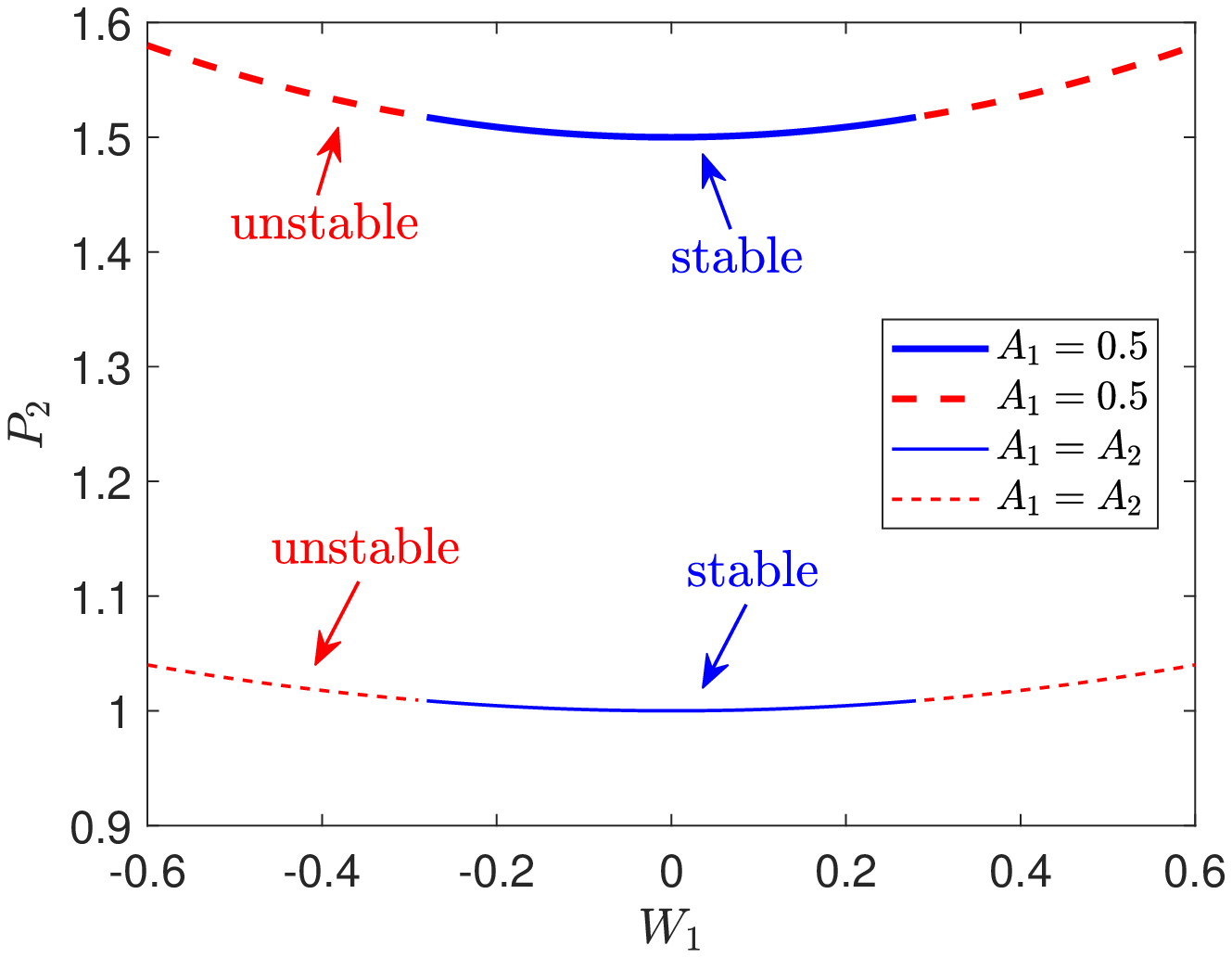}}
\subfigure[]{
\includegraphics[scale=0.57]{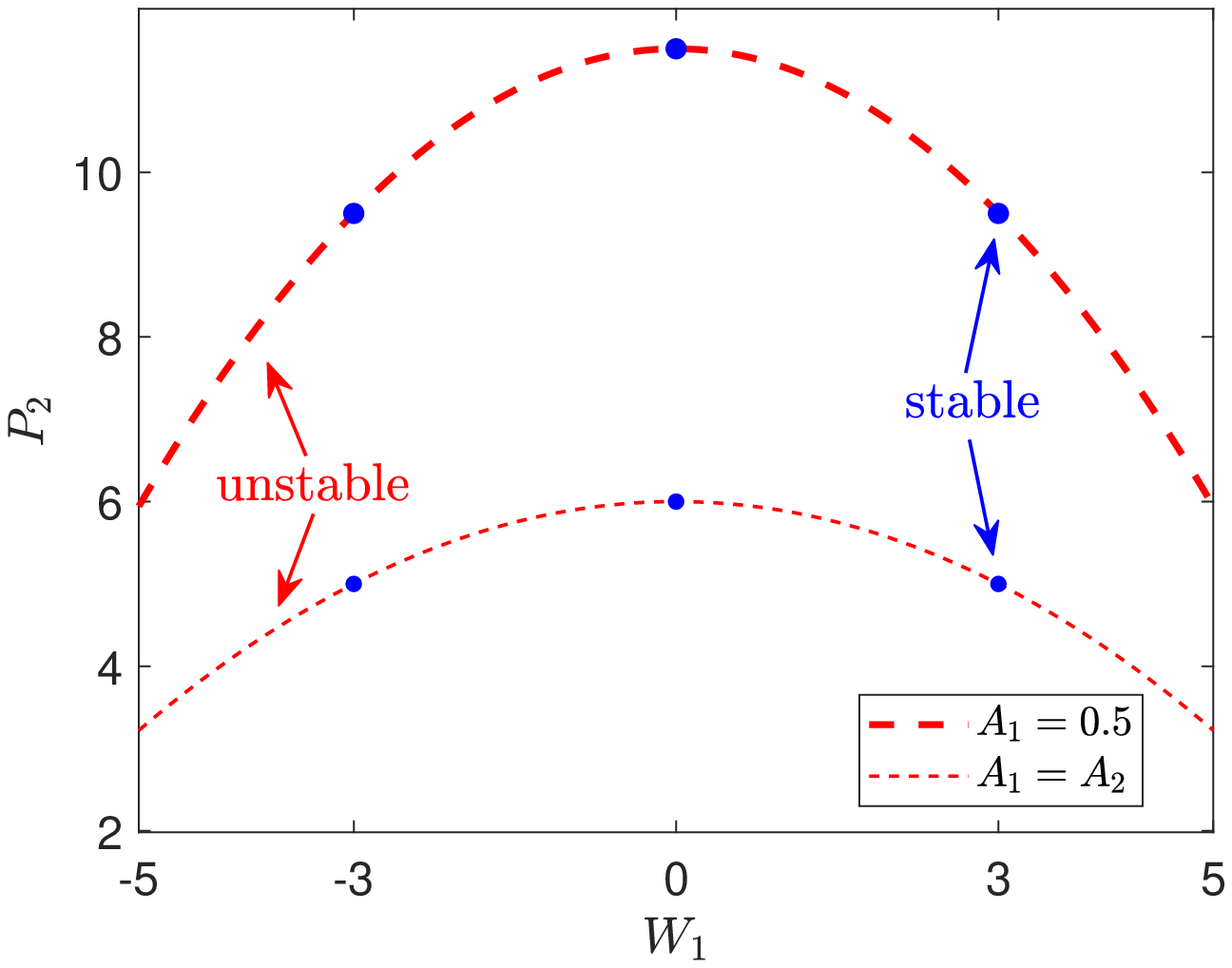}}
\end{center}
\figcaption{\footnotesize
The relationship between power of nonlinear mode $\phi_2$ and $W_1$. The parameters are chosen as: $W_1 = W_2$, and (a) $V_1 = V_2 = 1$, $a_1 = 1$; (b) $V_1 = V_2 = 8$, $a_1 = -1$.
}
\label{2}
\end{figure}

Since the above situations are obtained in the case of $A_1 = 0.5$, and $A_2$ is obtained by Eq.~\ref{jie1}. Next, we consider the case of $A_1 = A_2$. The relationships between $P_2$ and the parameter $W_1$ are shown in Figs.~\ref{2}. We can find that when other parameters are fixed, $P_2$ and $|W_1|$ are positively correlated in the focusing case, while they are negatively correlated in the defocusing case. In addition, for both cases $A_1 = 0.5$ and $A_1 = A_2$, the intervals of $W_1$ have no difference when the solutions are stable.

\begin{figure}[htbp]
\begin{center}
\renewcommand{\captionfont}{\scriptsize}
\renewcommand{\captionlabelfont}{\scriptsize}
\renewcommand{\figurename}{Figs.\,}
\subfigure[]{
\includegraphics[scale=0.37]{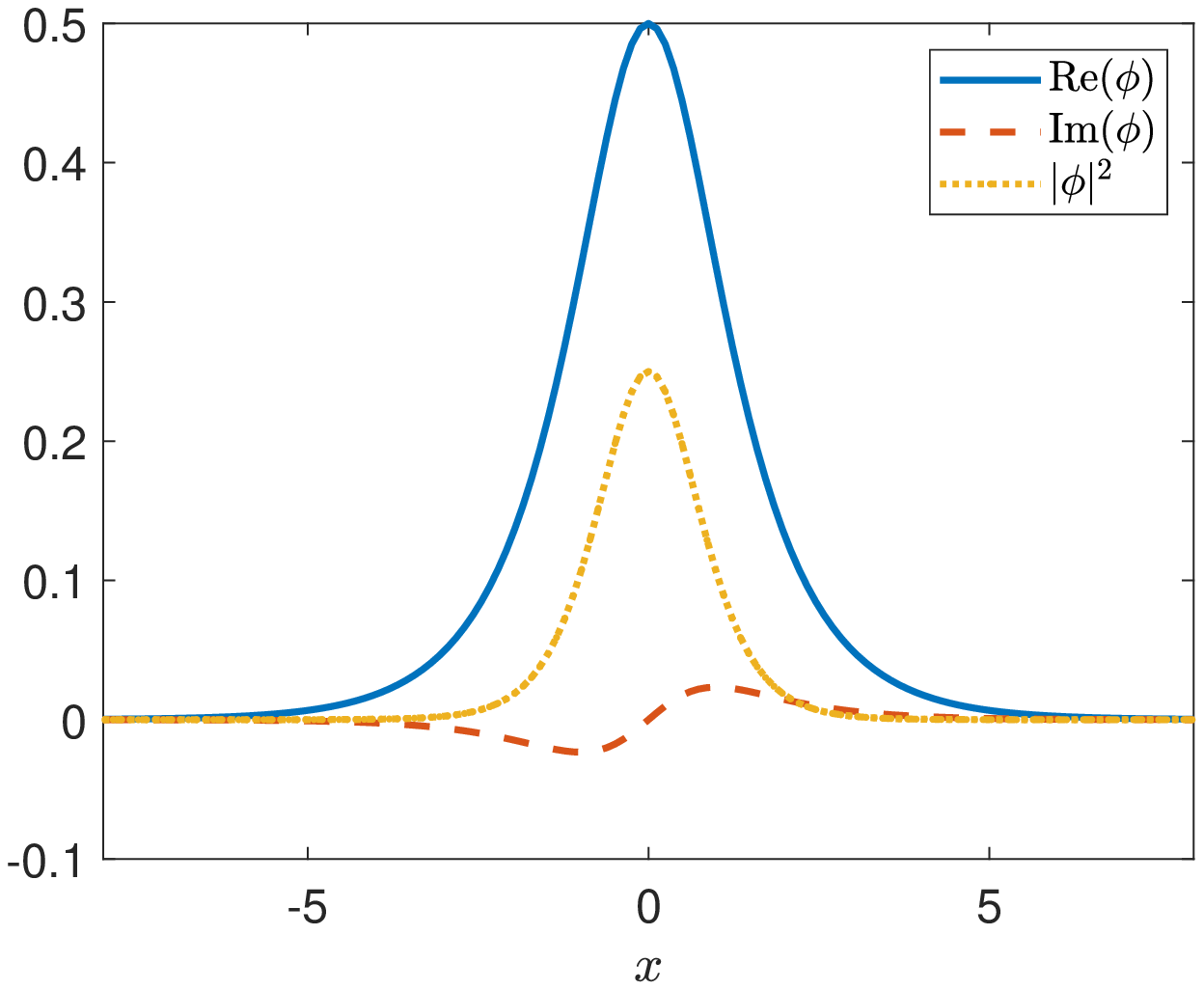}}
\subfigure[]{
\includegraphics[scale=0.37]{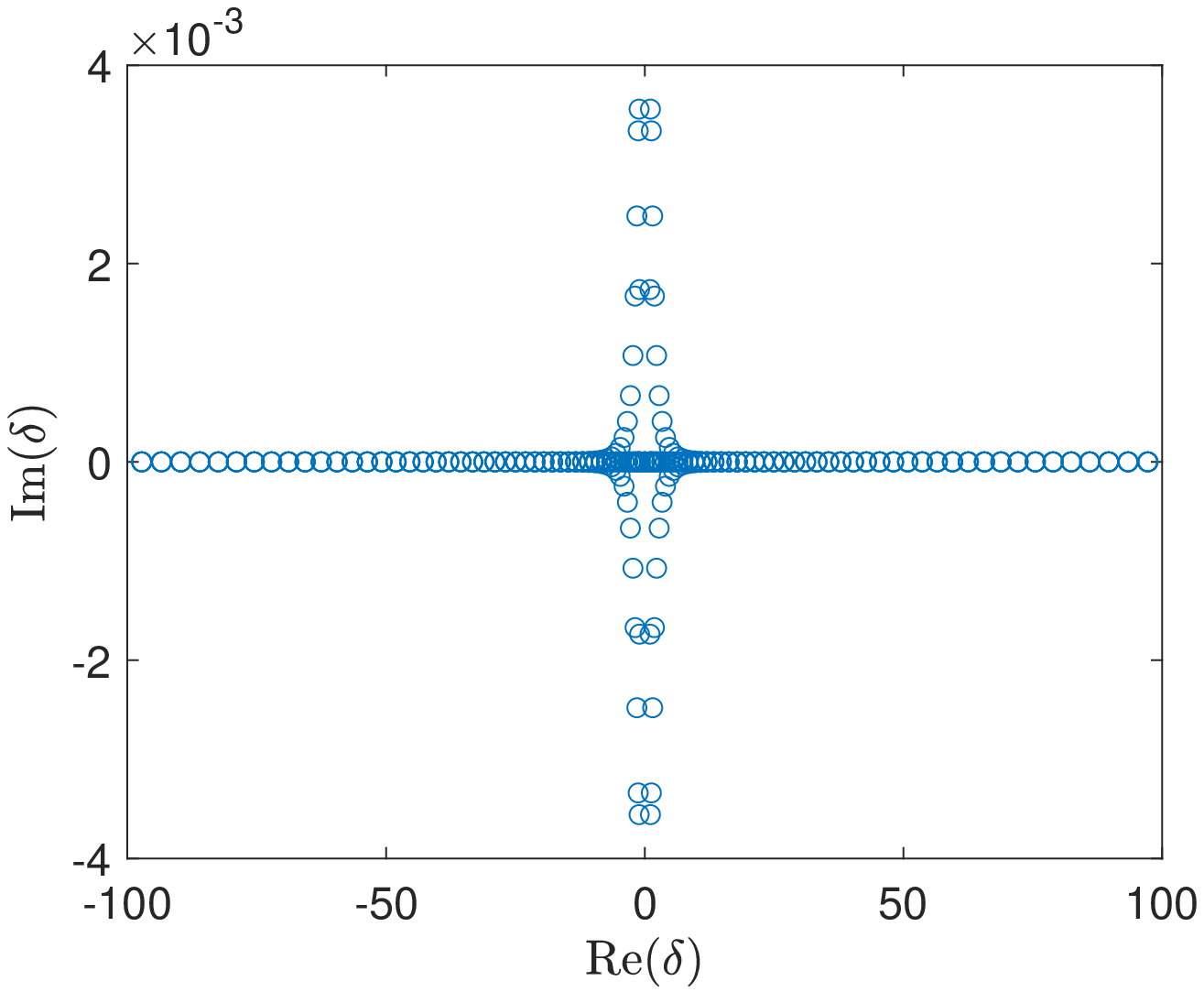}}
\subfigure[]{
\includegraphics[scale=0.37]{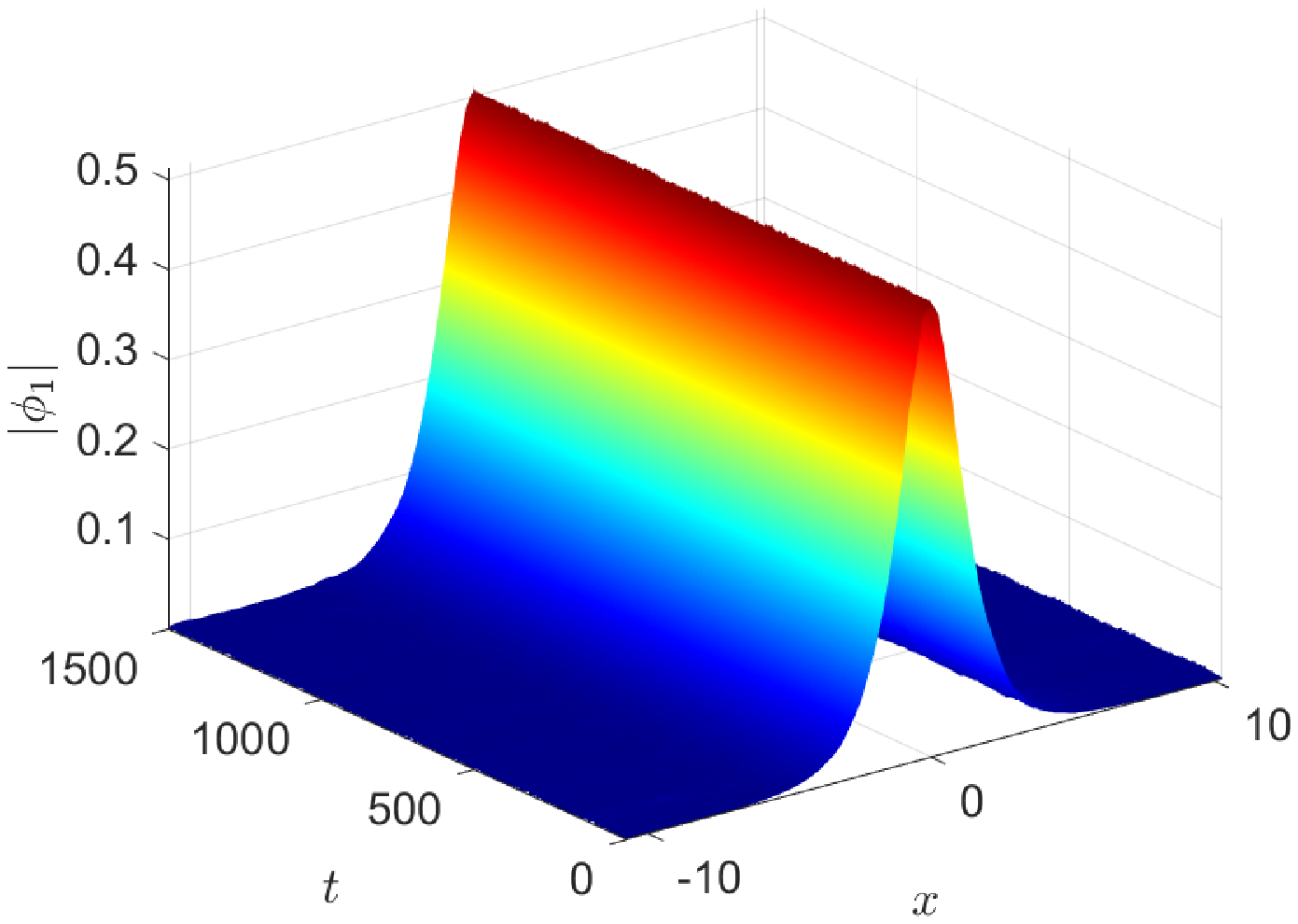}} \\
\subfigure[]{
\includegraphics[scale=0.37]{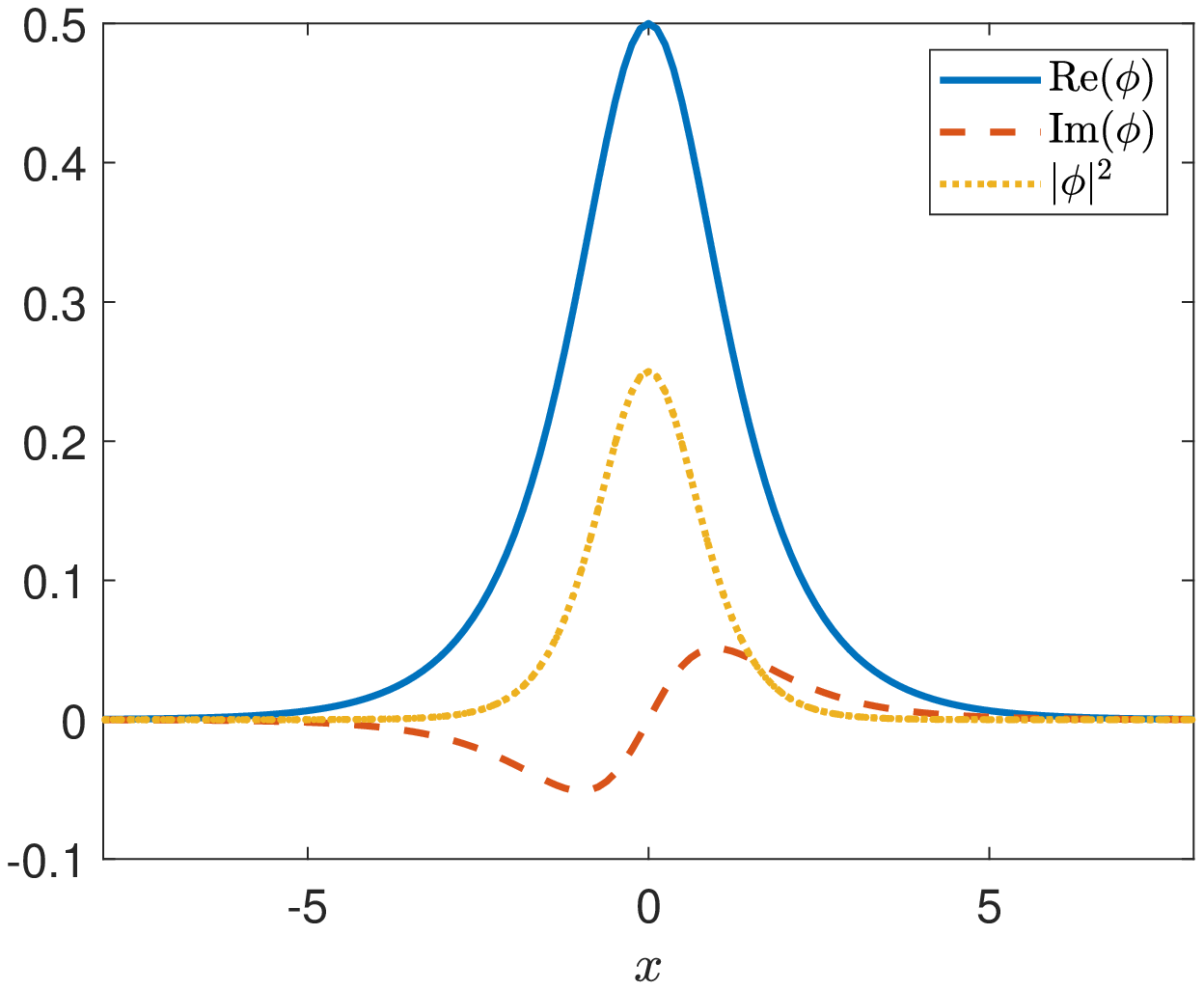}}
\subfigure[]{
\includegraphics[scale=0.37]{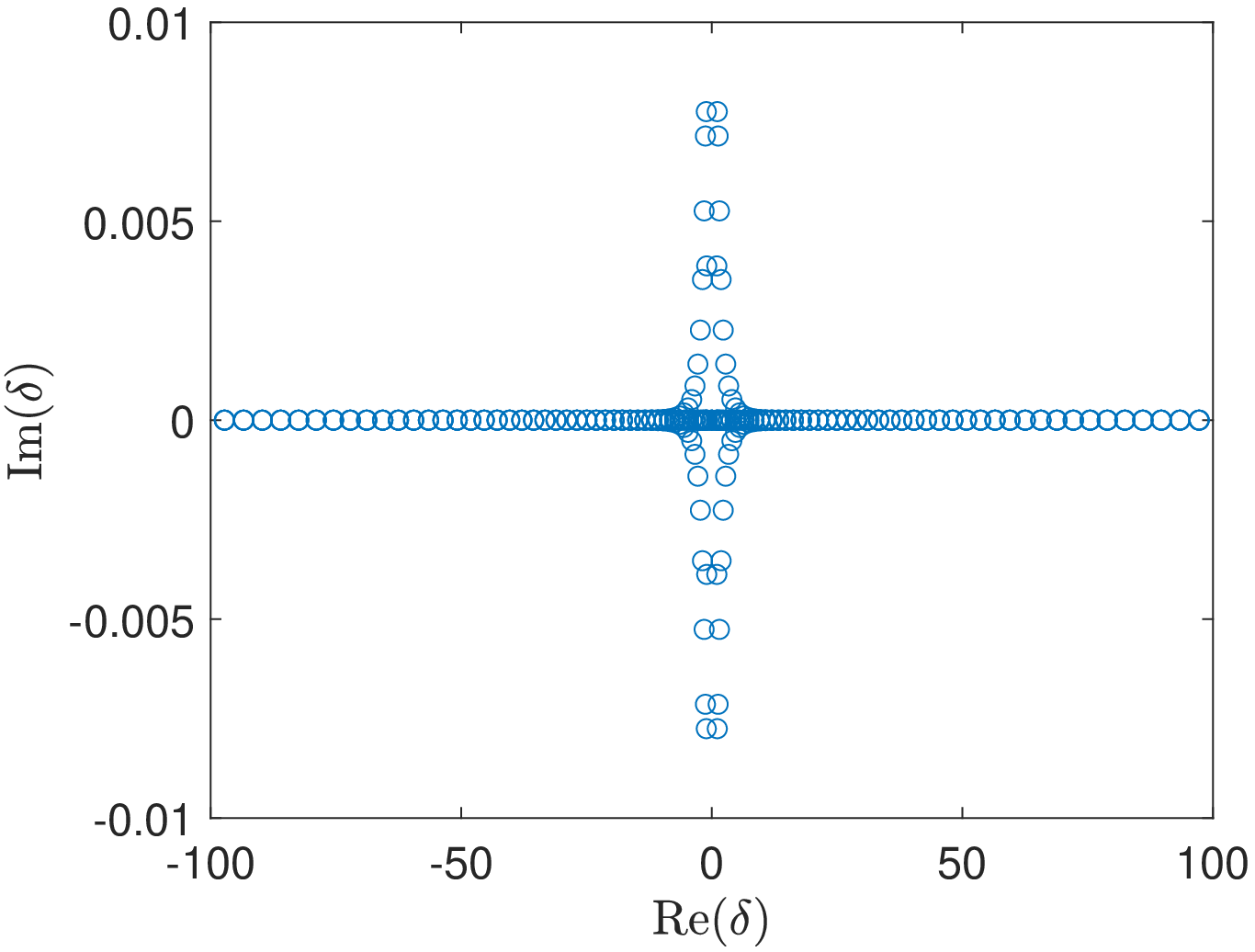}}
\subfigure[]{
\includegraphics[scale=0.37]{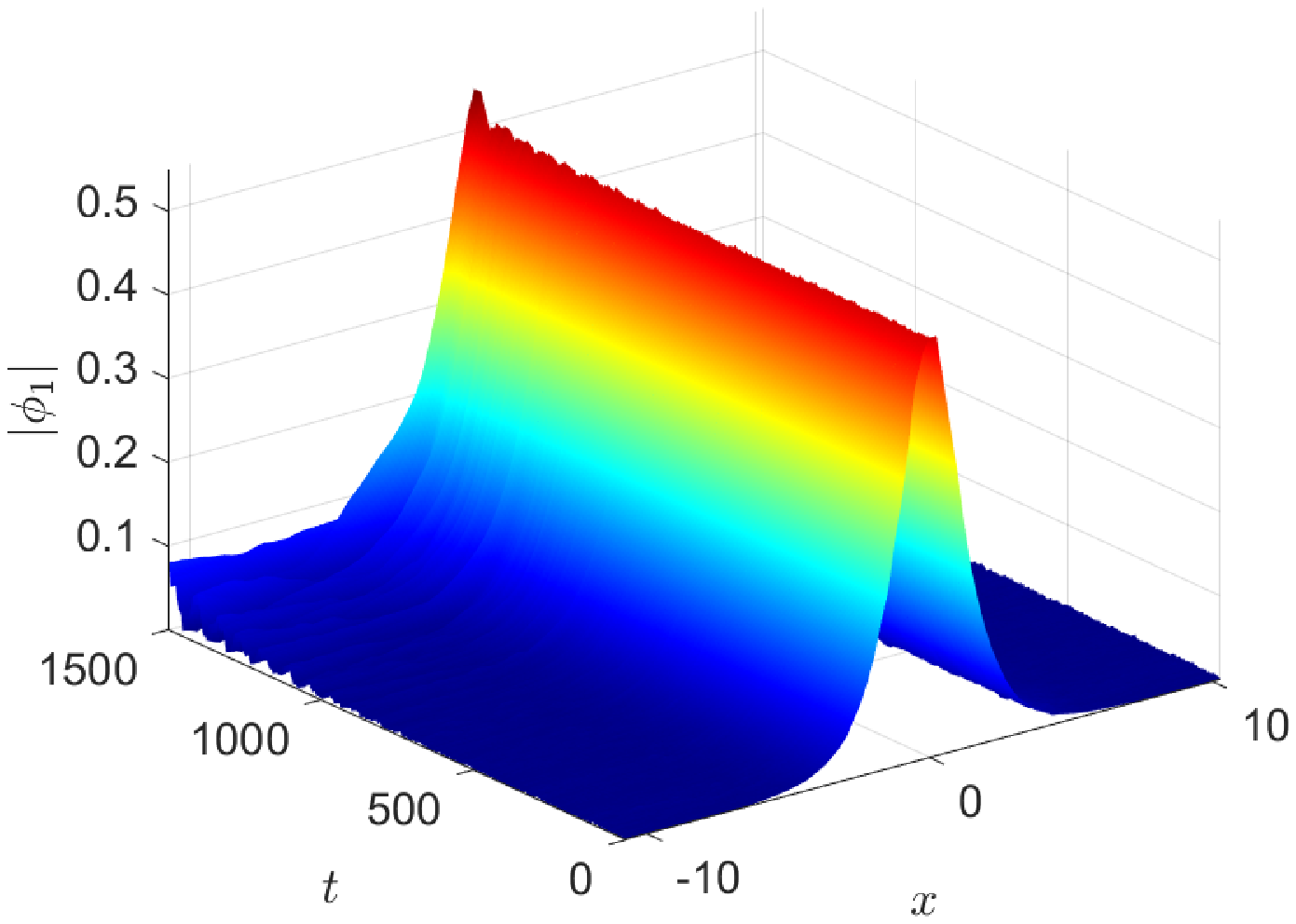}}
\end{center}
\figcaption{\footnotesize
(a, d) The soliton solutions. (b, e)  Linear stability eigenvalues. (c, f) Stable or unstable propagations of nonlinear modes. The parameters are chosen as: $a_1 = 1$, $A_1 = 0.5$, $V_1 = V_2 = 1$, and (a-c) $W_1 = W_2 = 0.25$; (d-f) $W_1 = W_2 = 0.55$.
}
\label{3}
\end{figure}

\begin{figure}[htbp]
\begin{center}
\renewcommand{\captionfont}{\scriptsize}
\renewcommand{\captionlabelfont}{\scriptsize}
\renewcommand{\figurename}{Figs.\,}
\subfigure[]{
\includegraphics[scale=0.37]{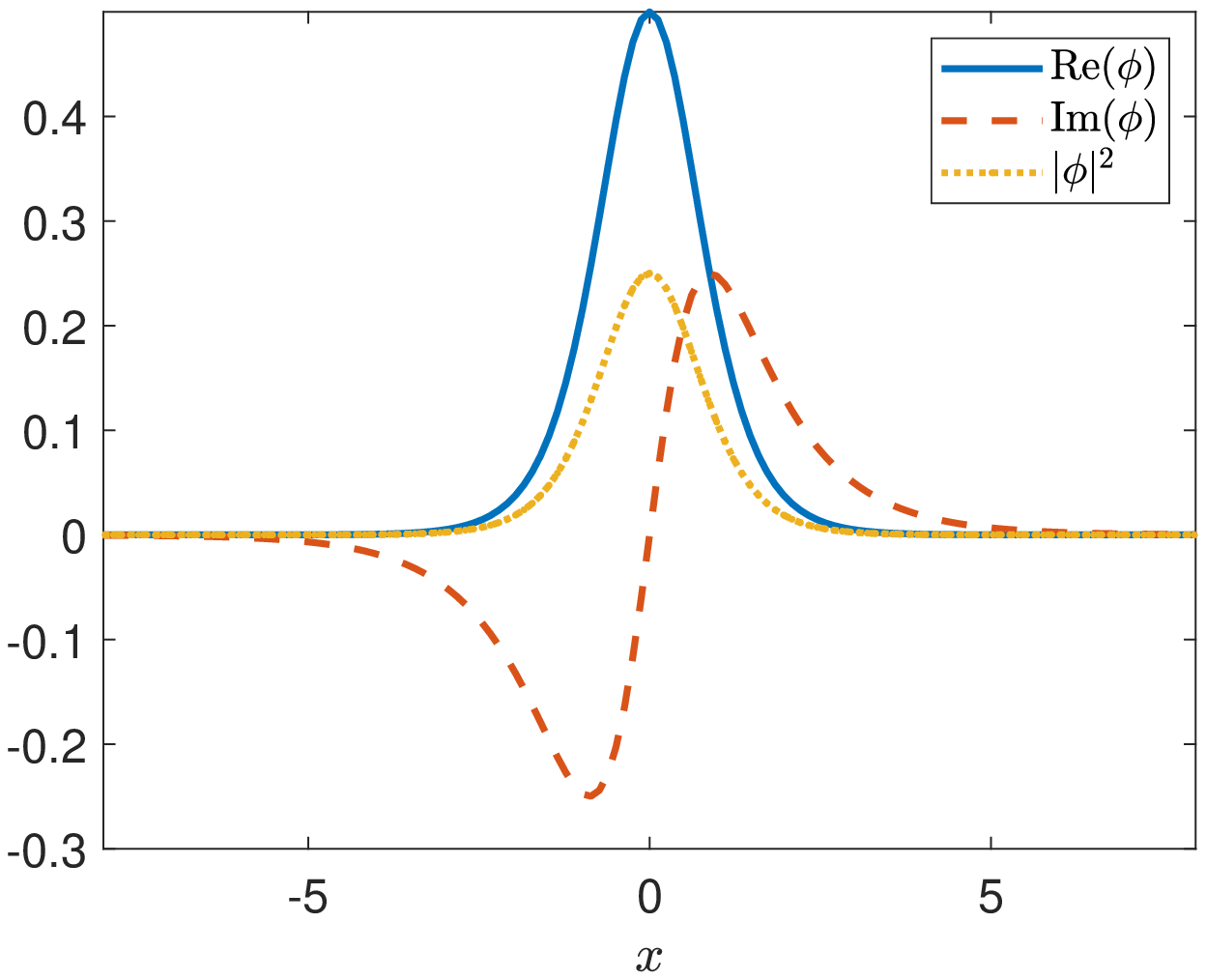}}
\subfigure[]{
\includegraphics[scale=0.37]{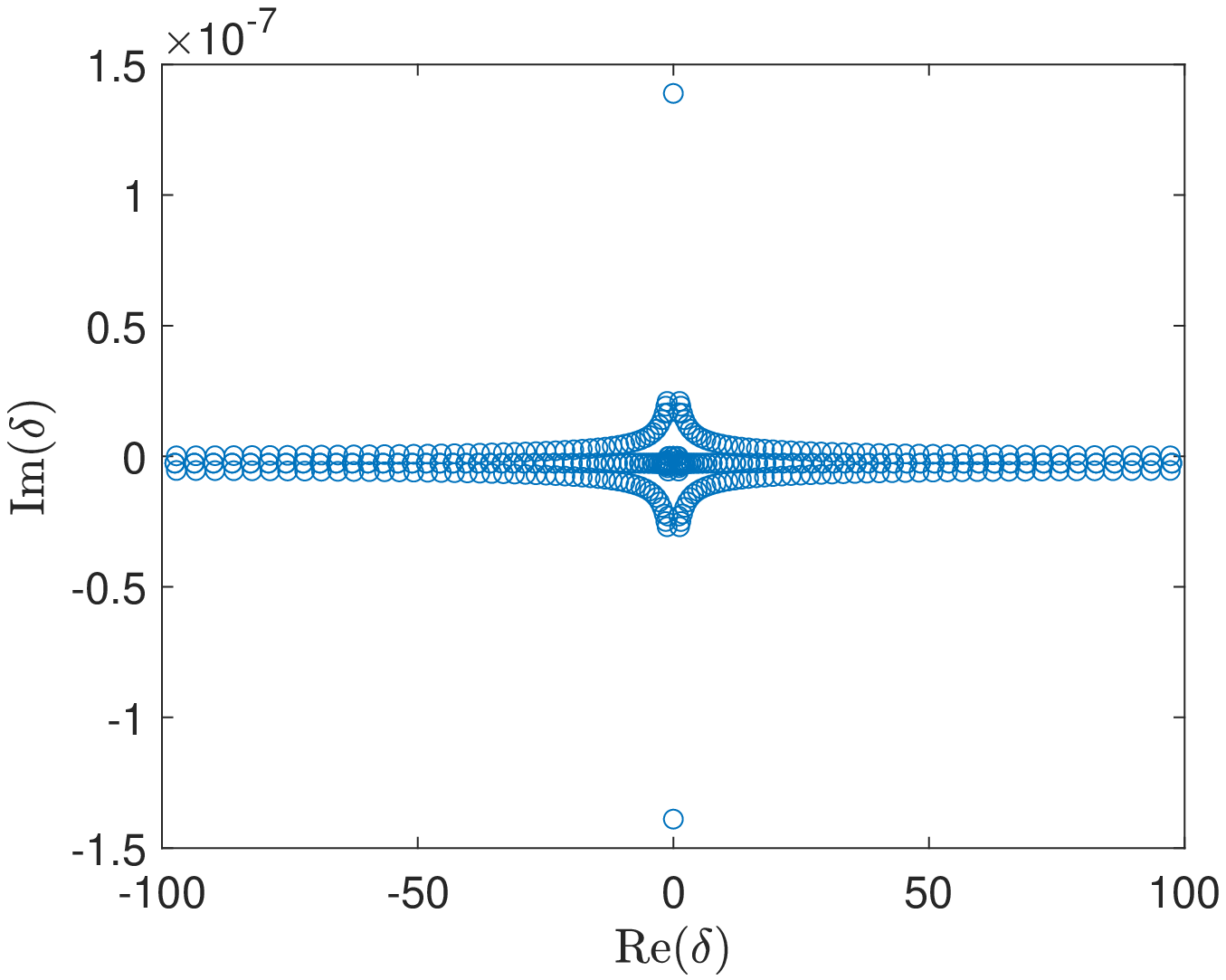}}
\subfigure[]{
\includegraphics[scale=0.37]{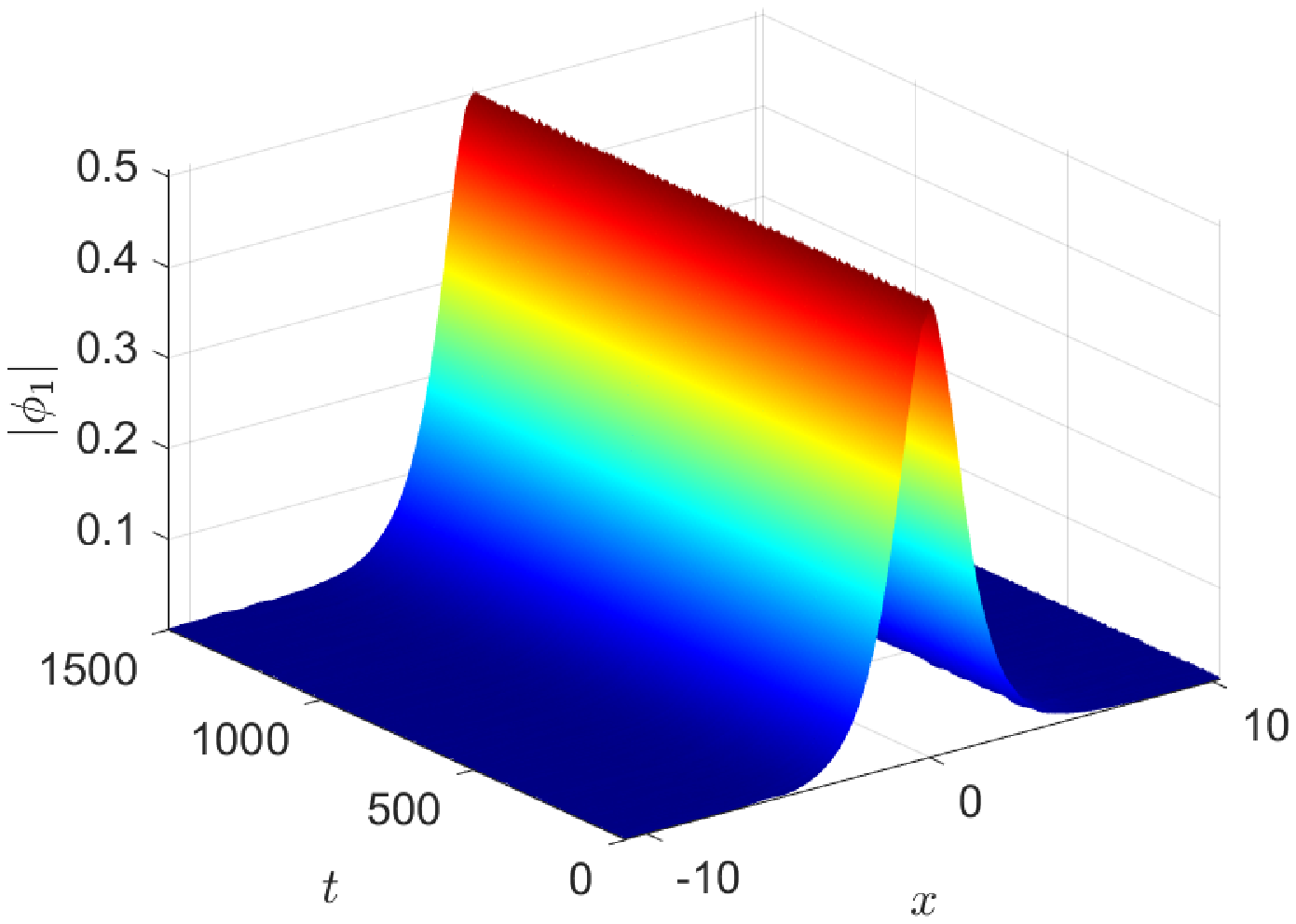}} \\
\subfigure[]{
\includegraphics[scale=0.37]{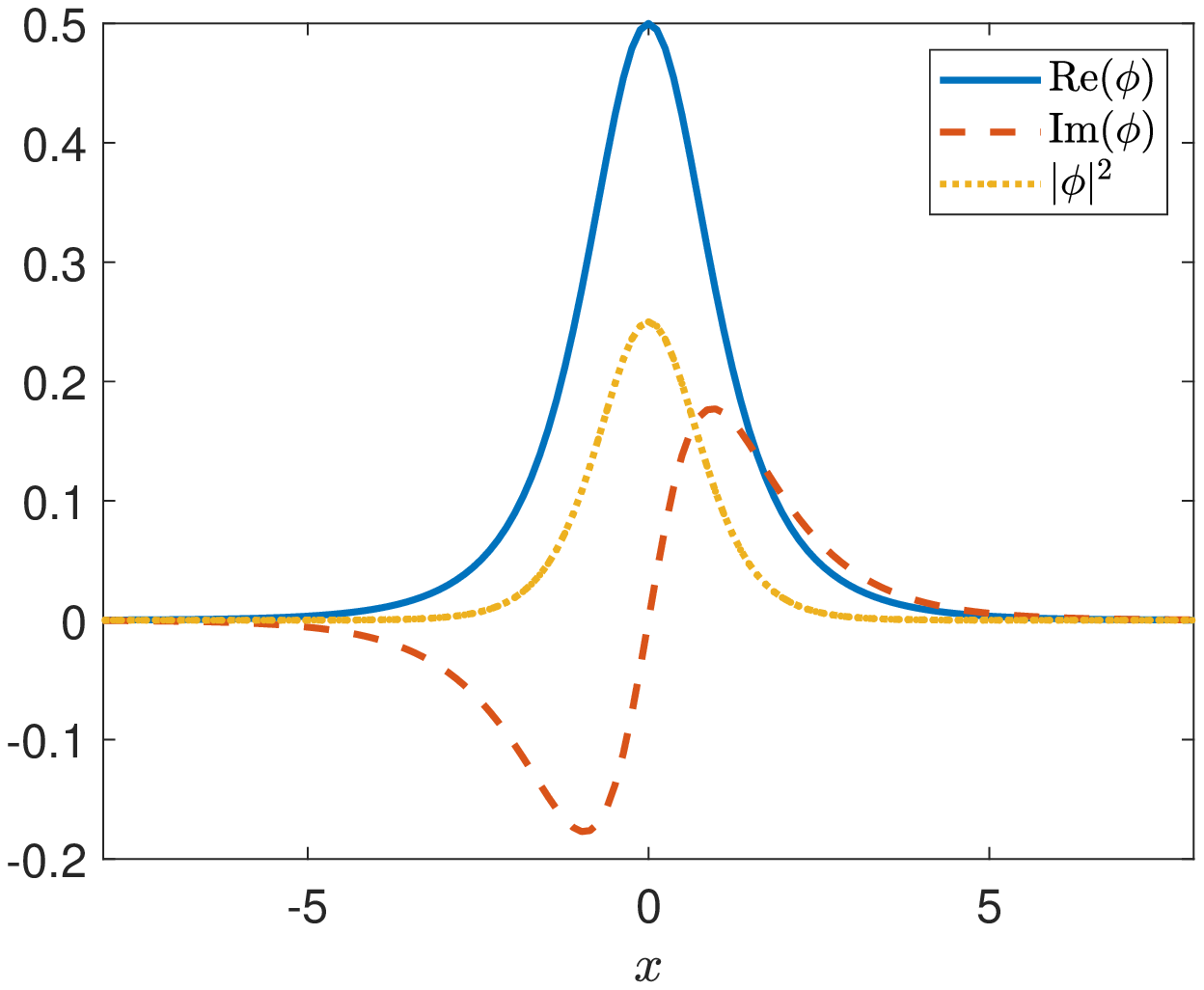}}
\subfigure[]{
\includegraphics[scale=0.37]{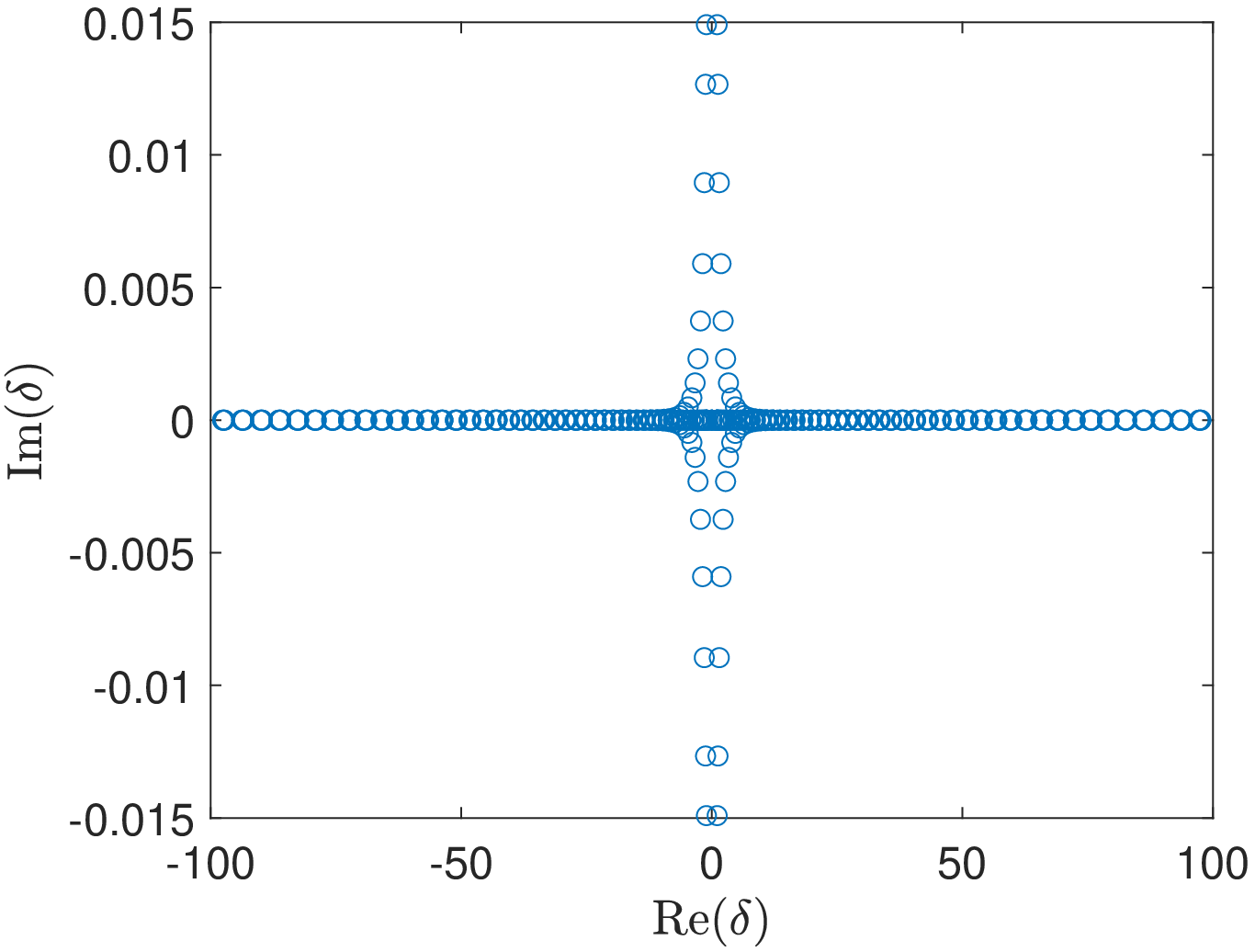}}
\subfigure[]{
\includegraphics[scale=0.37]{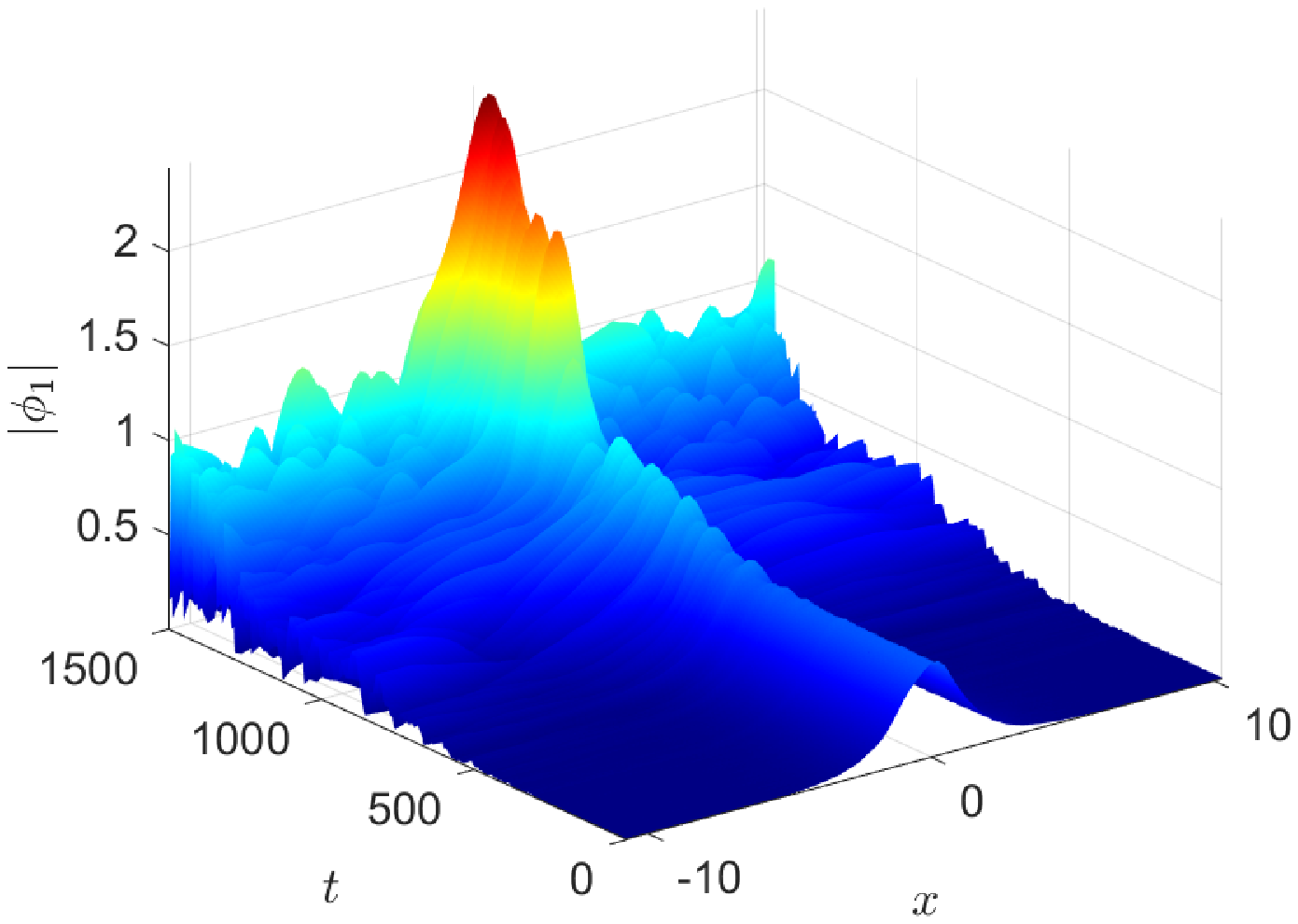}}
\end{center}
\figcaption{\footnotesize
(a, d) The soliton solutions. (b, e)  Linear stability eigenvalues. (c, f) Stable or unstable propagations of nonlinear modes. The parameters are chosen as: $a_1 = -1$, $A_1 = 0.5$, $V_1 = V_2 = 8$, and (a-c) $W_1 = W_2 = 3$; (d-f) $W_1 = W_2 = 2$.
}
\label{4}
\end{figure}

In particular, for the fixed parameters $a_1 = a_2 = 1$, $A_1 = 0.5$, $V_1 = V_2 = 1$, Figs.~\ref{3}(a)-\ref{3}(c) display the stable soliton for $W_1 = W_2 = 0.25$ while Figs.~\ref{3}(d)-\ref{3}(f) display the unstable soliton for $W_1 = W_2 = 0.55$; for the fixed parameters $a_1 = a_2 = -1$, $A_1 = 0.5$, $V_1 = V_2 = 8$, Figs.~\ref{4}(a)-\ref{4}(c) display the stable soliton for $W_1 = W_2 = 3$ while Figs.~\ref{4}(d)-\ref{4}(f) display the unstable soliton for $W_1 = W_2 = 2$.

\begin{figure}[htbp]
\begin{center}
\renewcommand{\captionfont}{\scriptsize}
\renewcommand{\captionlabelfont}{\scriptsize}
\renewcommand{\figurename}{Figs.\,}
\subfigure[]{
\includegraphics[scale=0.37]{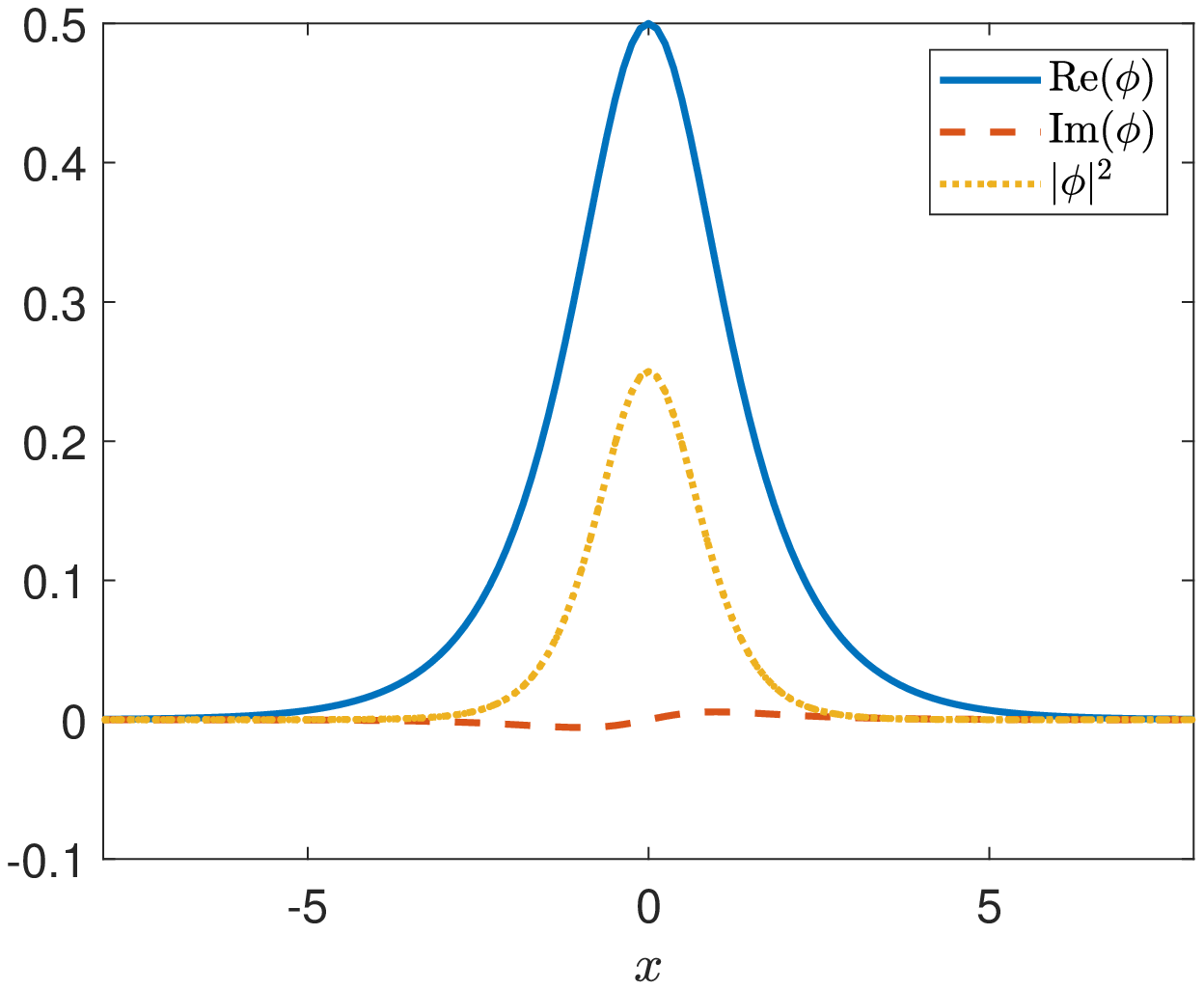}}
\subfigure[]{
\includegraphics[scale=0.37]{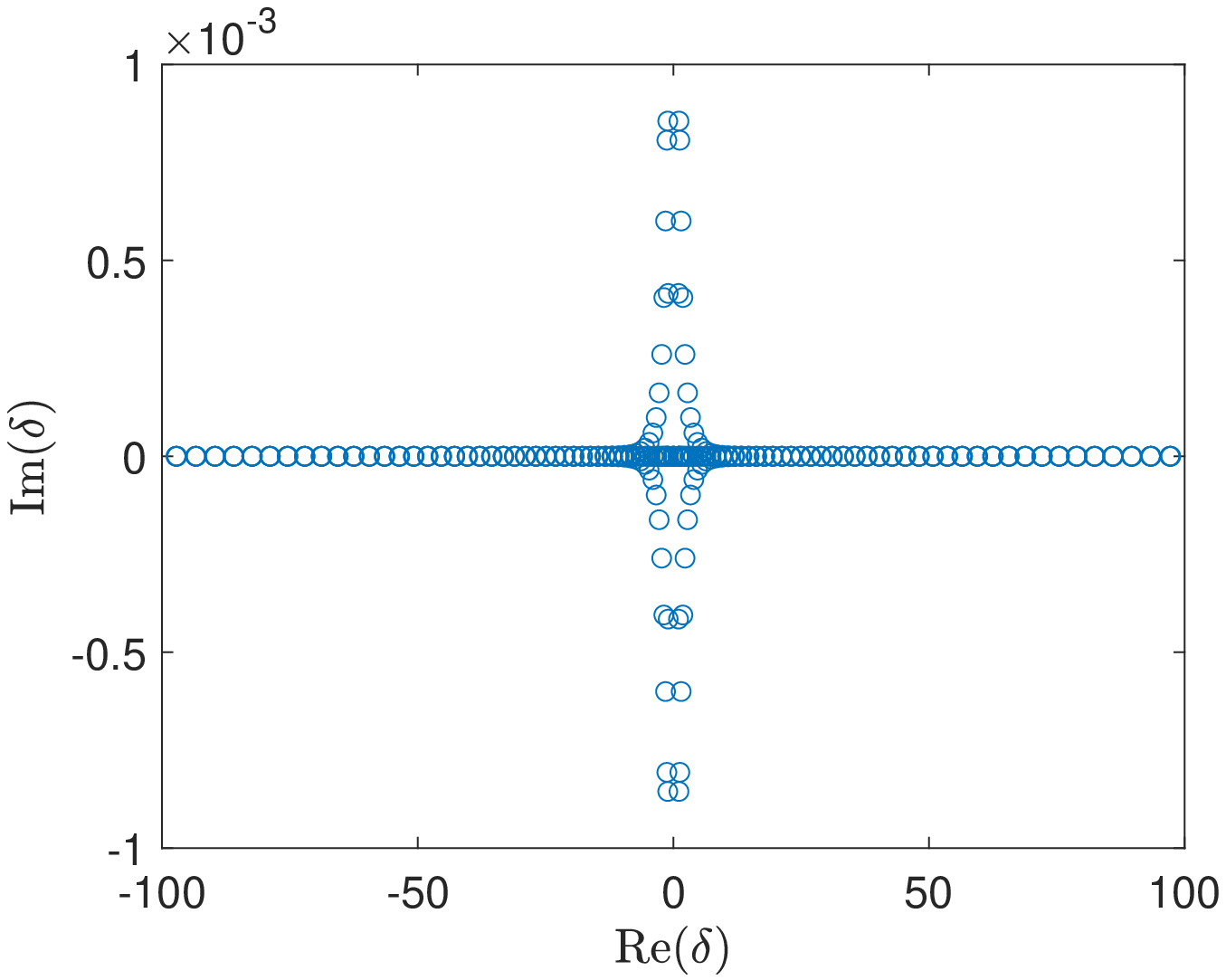}}
\subfigure[]{
\includegraphics[scale=0.37]{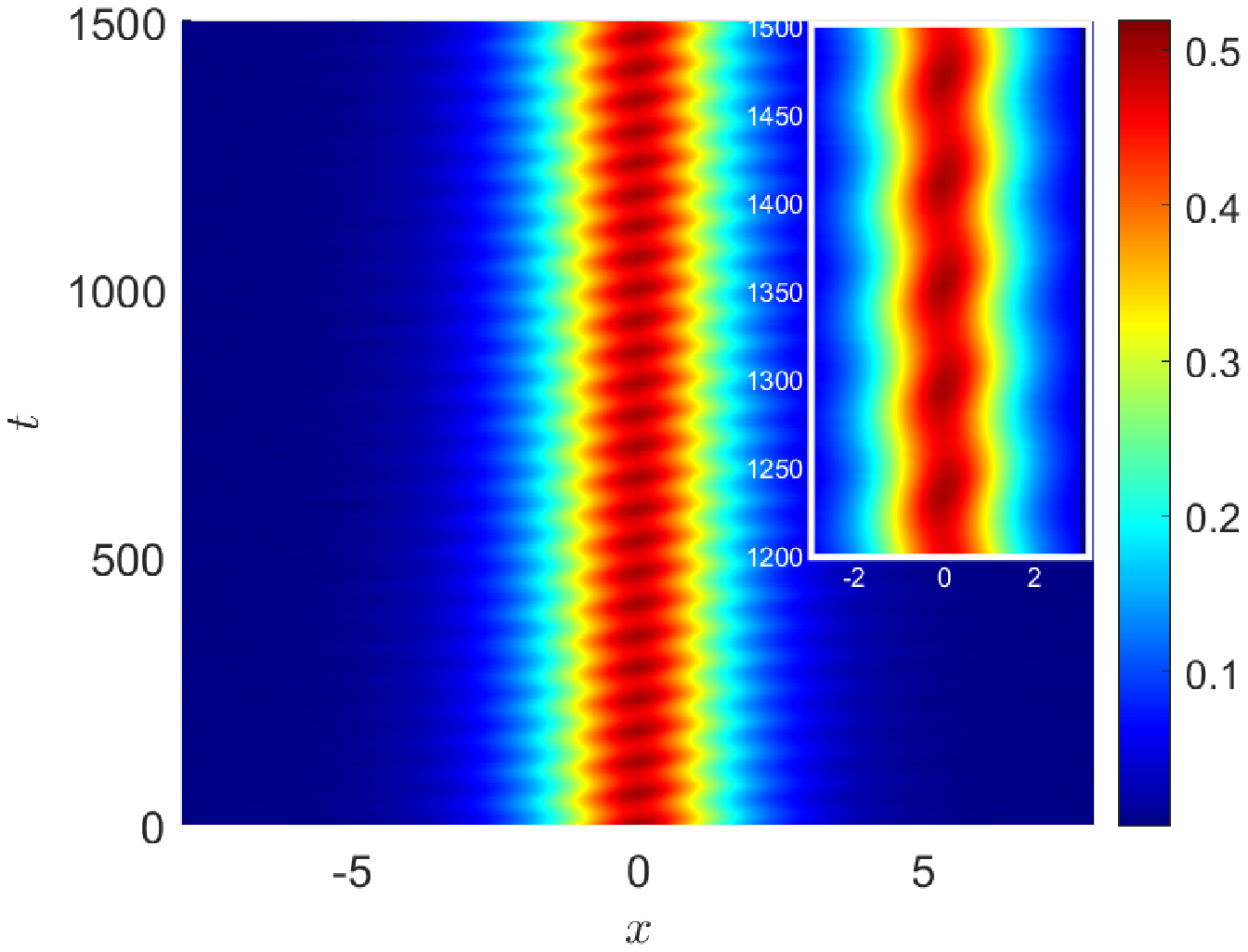}}
\end{center}
\figcaption{\footnotesize
(a) The soliton solutions. (b)  Linear stability eigenvalues. (c) Stable propagations of nonlinear modes. The parameters are chosen as: $a_1 = 1$, $A_1 = 0.5$, $V_1 = V_2 = 0.01$, $W_1 = W_2 = 0.06$.
}
\label{5}
\end{figure}

Furthermore, in the focusing case, the amplitude of the nonlinear mode is periodically oscillating when $V_1$ and $W_1$ are sufficiently small, and it experiences more than $5$ periods within $1200 \leq t \leq 1500$ [see Figs.~\ref{5}].

\section{Adiabatic excitation for the nonlinear modes}
\label{sec-4}

\begin{figure}[htbp]
\begin{center}
\renewcommand{\captionfont}{\scriptsize}
\renewcommand{\captionlabelfont}{\scriptsize}
\renewcommand{\figurename}{Figs.\,}
\subfigure[]{
\includegraphics[scale=0.37]{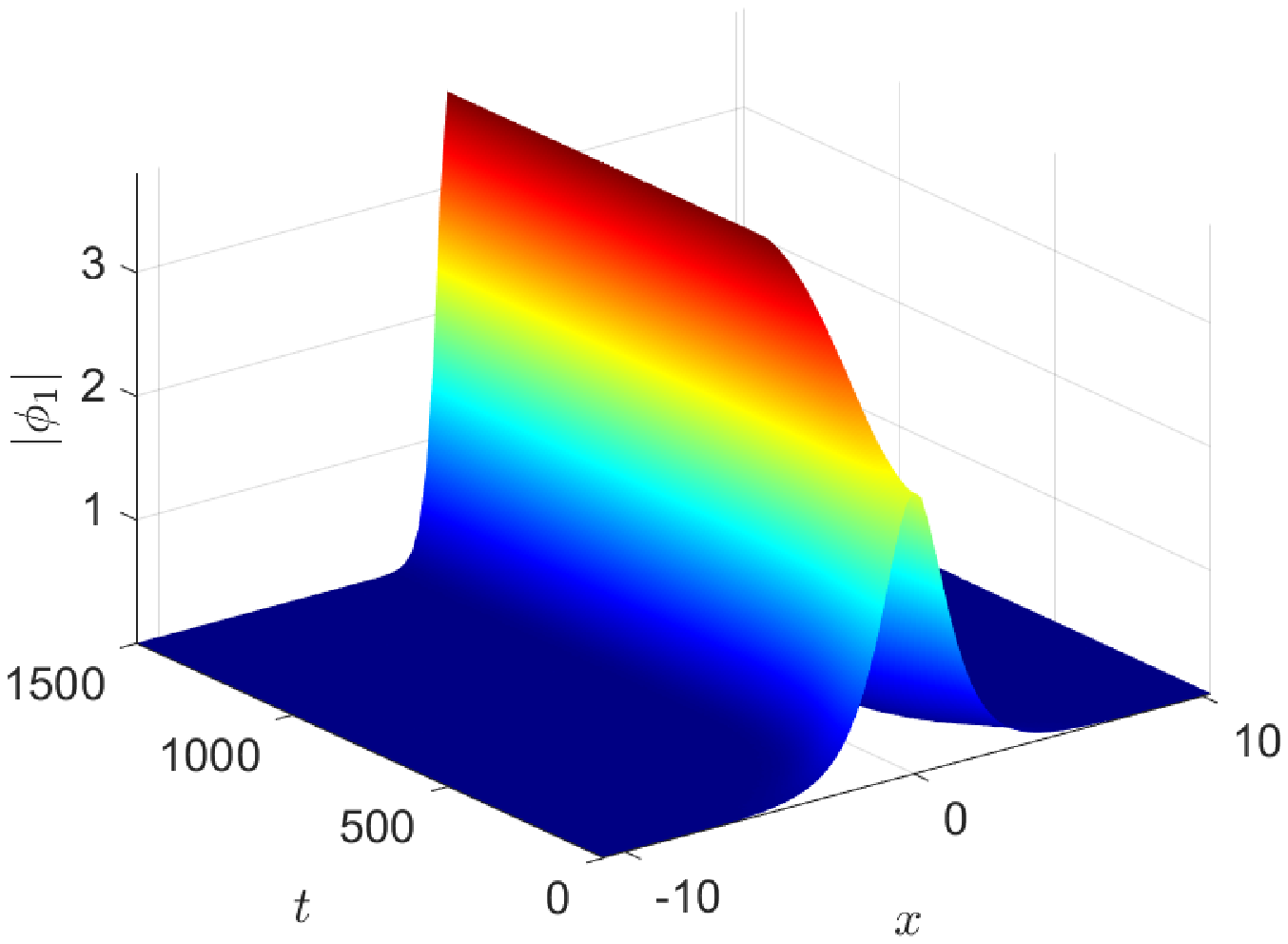}}
\subfigure[]{
\includegraphics[scale=0.37]{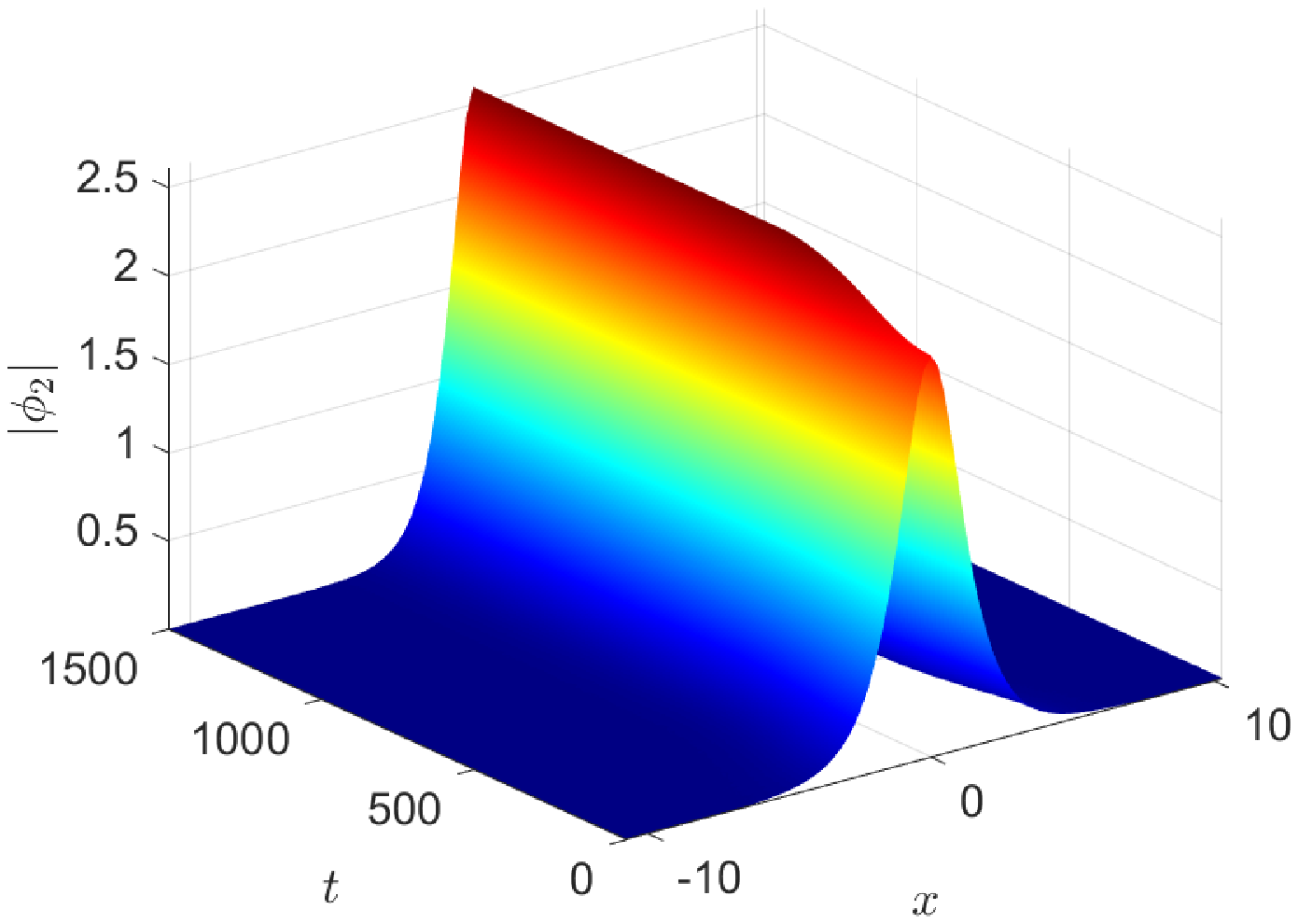}}
\subfigure[]{
\includegraphics[scale=0.37]{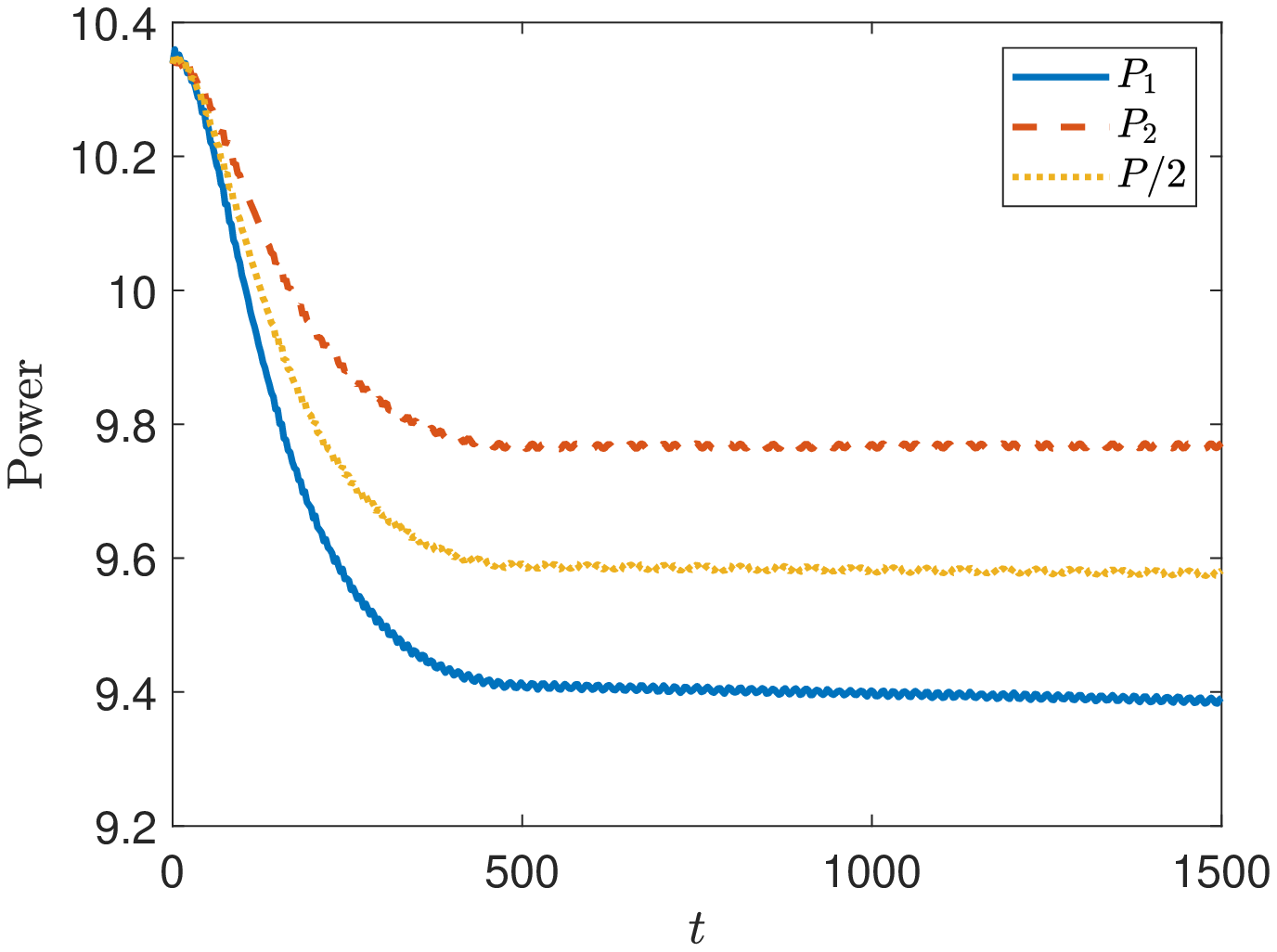}} \\
\subfigure[]{
\includegraphics[scale=0.37]{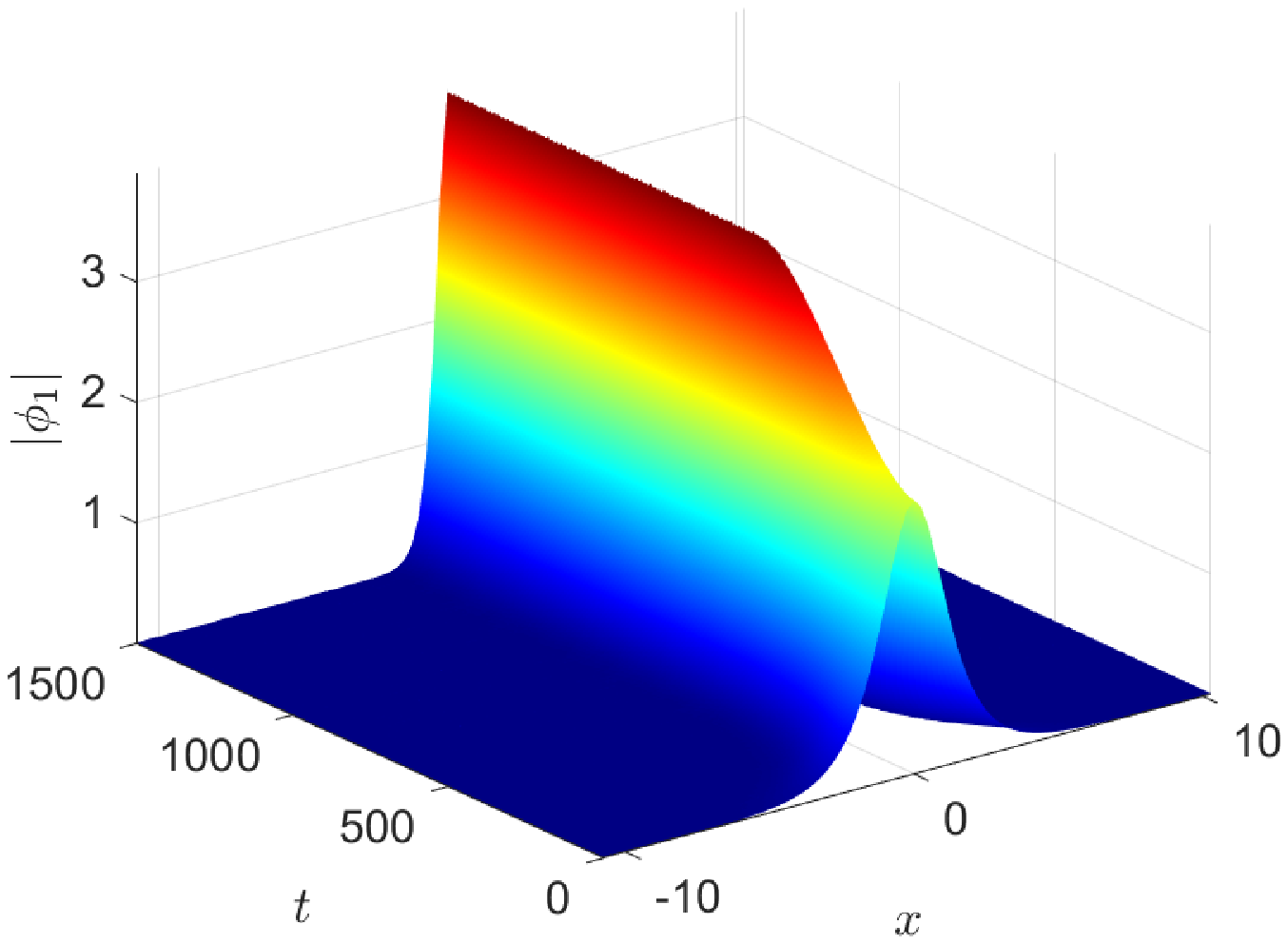}}
\subfigure[]{
\includegraphics[scale=0.37]{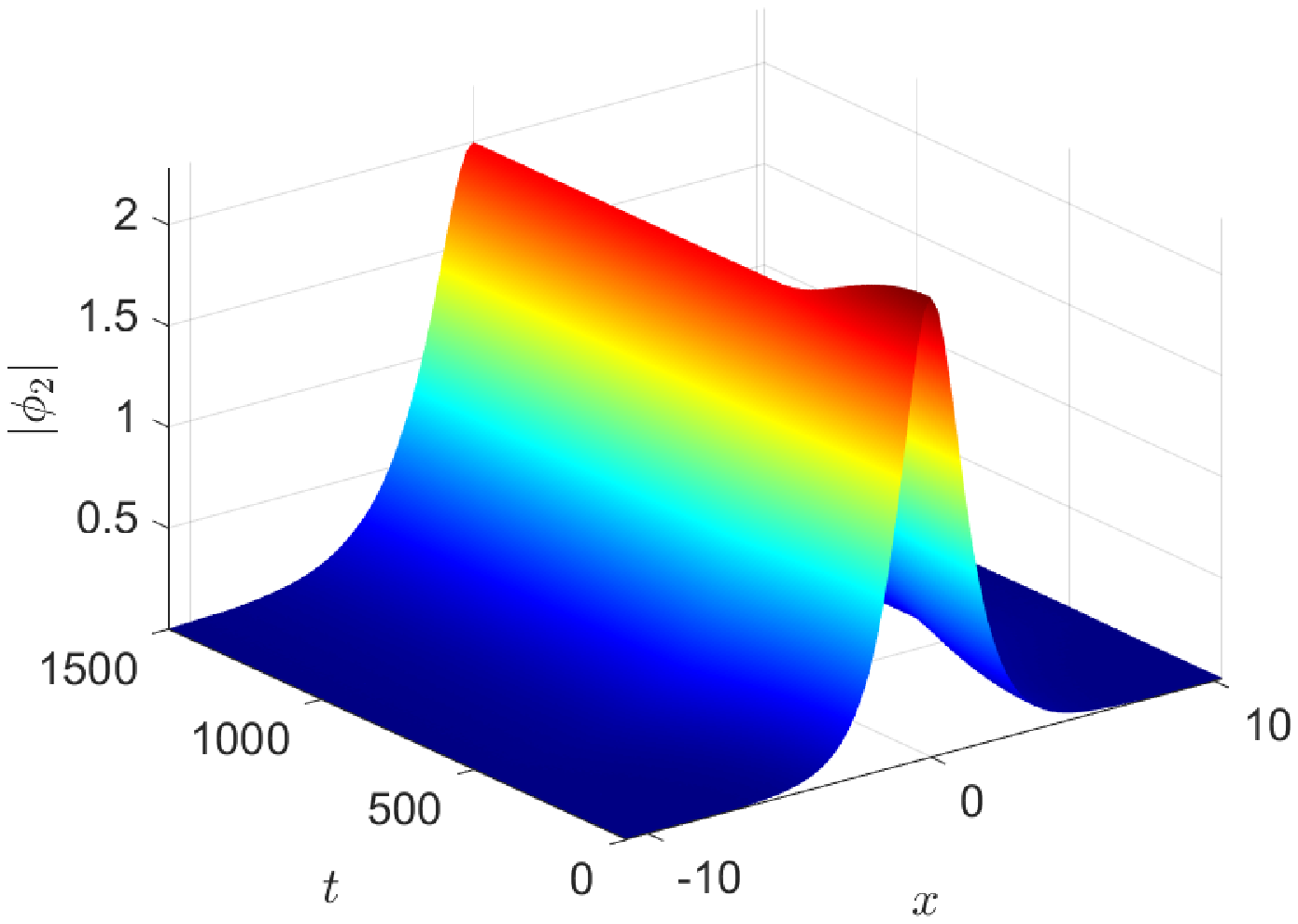}}
\subfigure[]{
\includegraphics[scale=0.37]{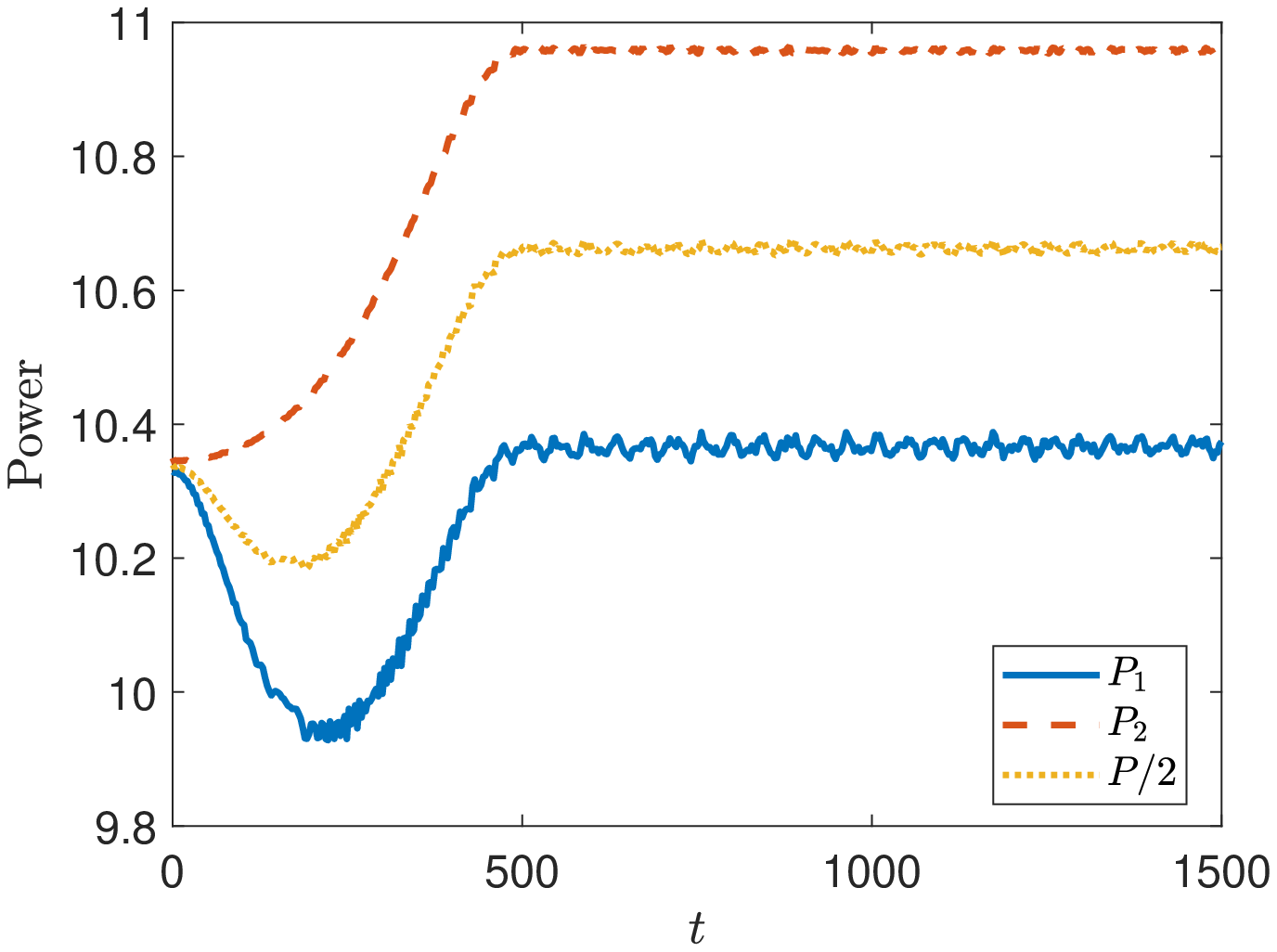}}
\end{center}
\figcaption{\footnotesize
Adiabatic excitation of nonlinear mode and its evolution. The parameters are chosen as: $A_1 = A_2 =2.2733$, $W_1 = W_2 = 0.55$, $V_1^{(\text{ini})} = 1$, $V_1^{(\text{end})} = 2$, $a_1^{(\text{ini})} = 0.1$, $a_1^{(\text{end})} = 1$ and (a-c) $V_2^{(\text{ini})} = 1$, $V_2^{(\text{end})} = 2$, $a_2 = 0.1$; (d-f) $V_2^{(\text{ini})} = 2$, $V_2^{(\text{end})} = 1$, $a_2 = 0.0033$.
}
\label{6}
\end{figure}

In this section, we consider excitations of the above-mentioned solitons via adiabatical changes of system parameters. We change the parameters as the functions of $t$. To modulate the system parameters smoothly, we consider the following ``switch-on" function:
\begin{eqnarray}\label{et1}
\zeta(t)=\left\{
\begin{array}{ll}
\zeta^{(\text{ini})}, & t = 0, \\[8pt]
\dfrac{\zeta^{(\text{end})}-\zeta^{(\text{ini})}}{2}\left[1+\sin\left(\dfrac{\pi t}{500}-\dfrac{\pi}{2}\right)\right]+\zeta^{(\text{ini})}, & 0 < t < 500, \\[8pt]
\zeta^{(\text{end})}, & 500 \leq t \leq 1500,
\end{array}
\right.
\end{eqnarray}
where $\zeta^{(\text{ini})}$, $\zeta^{(\text{end})}$ respectively represent the real initial-state and final-state parameters~\cite{yan2015,Shen2018}. Adiabatic excitation includes two stages: excitation stage $(0 < t < 500)$ and propagation stage $(500 \leq t \leq 1500)$. We consider two cases of excitations by setting $a_1$, $V_1$ and $V_2$ to be functions of $t$, that is $a_1 \rightarrow a_1(t)$, $V_1 \rightarrow V_1(t)$ and $V_2 \rightarrow V_2(t)$. To facilitate the display of power changes over time, we set $A_1 = A_2$, $W_1 = W_2 = 0.55$, $V_1^{(\text{ini})} = 1$, $V_1^{(\text{end})} = 2$, $a_1^{(\text{ini})} = 0.1$, $a_1^{(\text{end})} = 1$. Firstly, we set $V_2^{(\text{ini})} = 1$, $V_2^{(\text{end})} = 2$, $a_2 = 0.1$, and the power of nonlinear modes is reduced [see Figs.~\ref{6}a-\ref{6}c]. Then, we set $V_2^{(\text{ini})} = 2$, $V_2^{(\text{end})} = 1$, $a_2 = 0.0033$, and the total power of nonlinear modes is to decrease and then increase [see Figs.~\ref{6}d-\ref{6}f]. The above results mean that the power is not conserved during adiabatic excitations, and it has a correlation with the initial and final state potential parameters.

\section{conclusion}
\label{sec-5}

In conclusion, we study the nonlinear modes in two-component Bose-Einstein condensates with $\PT$-symmetric Scarf-II potential, which can be described by the coupled Gross-Pitaevskii equations. We investigate the linear stability of the nonlinear modes and validate the results by evolving them with $5\%$ perturbations as an initial condition. We find that solitons tend to be unstable with the increase of $|W_1|$ in the focusing case. It is worth noting that when $|W_1|=3$, solitons are stable in the defocusing case. In the focusing case, the amplitude of the nonlinear mode is periodically oscillating  when $V_1$ and $W_1$ are sufficiently small. Finally, we consider excitations of the solitons via adiabatical changes of system parameters, then we find that the power is not conserved during this adiabatic excitation. These findings of nonlinear modes can be potentially applied to physical experiments of matter waves in Bose-Einstein condensates.

In addition, we can consider other $\PT$-symmetric potentials in the coupled Gross-Pitaevskii equations. Due to the limitations of the parameters in this model, the amplitudes of the two solutions are constrained by a certain relationship. Therefore, we can also consider the case of unequal intra-component and inter-component interactions.

\vspace{7mm} \noindent {\large{\bf Acknowledgements}}

We express our sincere thanks to the editor, referees and all the members of our discussion group for their valuable comments. This work was supported by the National Training Program of Innovation (Grant numbers 202210019045). The funding body plays an important role in the design of the study, in analysis, calculation, and in writing of the manuscript.

\vspace{3mm} \noindent {\large{\bf Declarations of interest}}

none.

\vspace{3mm} \noindent {\large{\bf Funding}}

This work was supported by the National Training Program of Innovation (Grant numbers 202210019045). The funding body plays an important role in the design of the study, in analysis, calculation, and in writing of the manuscript.

\vspace{3mm} \noindent {\large{\bf References}}



\begin{thebibliography}{99}
\addtolength{\itemsep}{-0.8 em}
\begin{small}
\bibitem{Einstein1924}
A. Einstein,
Quantentheorie des einatomigen idealen gases,
{\em Sitz. Ber. Kgl. Preuss. Akad. Wiss.} (1924) 261-7.

\bibitem{Einstein1925}
A. Einstein,
Quantentheorie des einatomigen idealen gases,
{\em Sitzungber Preuss} {\bf 9} (1925) 3-14.


\bibitem{Anderson1995}
M. H. Anderson, J. R. Ensher, M. R. Matthews, C. E. Wieman, E. A. Cornell,
Observation of Bose-Einstein condensation in a dilute atomic vapor,
{\em Science} {\bf 269} (1995) 198-201.

\bibitem{Davis1995}
K. B. Davis, M. O. Mewes, M. R. Andrews, N. J. Van Druten, D. S. Durfee, D. M. Kurn, W. Ketterle,
Bose-Einstein condensation in a gas of sodium atoms,
{\em Phys. Rev. Lett.} {\bf 75} (1995) 3969-3973.

\bibitem{Strecker2002}
K. E. Strecker, G. B. Partridge, A. G. Truscott, and R. G. Hulet,
Formation and propagation of matter-wave soliton trains,
{\em Nature} {\bf 417} (2002) 150-153.


\bibitem{Myatt1997}
C. J. Myatt, E. A. Burt, R. W. Ghrist, E. A. Cornell, C. E. Wieman,
Production of Two Overlapping Bose-Einstein Condensates by Sympathetic Cooling,
{\em Phys. Rev. Lett.} {\bf 78} (1997) 586.

\bibitem{Pu1998}
H. Pu, N. P. Bigelow,
Properties of Two-Species Bose Condensates,
{\em Phys. Rev. Lett.} {\bf 80} (1998) 1130.

\bibitem{Modugno2001}
G. Modugno, G. Ferrari, G. Roati, R. J. Brecha, A. Simoni, M. Inguscio,
Bose-Einstein condensation of potassium atoms by sympathetic cooling,
{\em Science} {\bf 294} (2001) 1320.

\bibitem{Alon2005}
O. E. Alon, A. I. Streltsov, L. S. Cederbaum,
Zoo of Quantum Phases and Excitations of Cold Bosonic Atoms in Optical Lattices,
{\em Phys. Rev. Lett.} {\bf 95} (2005) 030405.

\bibitem{Kapale2005}
K. T. Kapale, J. P. Dowling,
Vortex Phase Qubit: Generating Arbitrary, Counterrotating, Coherent Superpositions in Bose-Einstein Condensates via Optical Angular Momentum Beams,
{\em Phys. Rev. Lett.} {\bf 95} (2005) 173601.

\bibitem{Bhongale2008}
S. G. Bhongale and E. Timmermans,
Phase Separated BEC for High-Sensitivity Force Measurement,
{\em Phys. Rev. Lett.} {\bf 100} (2008) 185301.

\bibitem{Zawadzki2010}
M. E. Zawadzki, P. F. Griffin, E. Riis, A. S. Arnold,
Spatial interference from well-separated split condensates,
{\em Phys. Rev. A} {\bf 81} (2010) 043608.

\bibitem{Zoest2010}
T. van Zoest et al.,
Bose-Einstein Condensation in Microgravity,
{\em Science} {\bf 328} (2010) 1540.


\bibitem{Victor2003}
V\'{i}ctor M. P\'{e}rez-Garc\'{i}a, Vadym Vekslerchik,
Soliton molecules in trapped vector nonlinear Schr\"{o}dinger systems,
{\em Phys. Rev. E} {\bf 67} (2003) 061804.

\bibitem{Gaspar2004}
G. D. Montesinos, V\'{i}ctor M. P\'{e}rez-Garc\'{i}a, Humberto Michinel,
Stabilized Two-Dimensional Vector Solitons,
{\em Phys. Rev. Lett.} {\bf 92} (2004) 133901.

\bibitem{Kasamatsu2004}
K. Kasamatsu, M. Tsubota,
Multiple Domain Formation Induced by Modulation Instability in Two-Component Bose-Einstein Condensates,
{\em Phys. Rev. Lett.} {\bf 93} (2004) 100402.

\bibitem{Malomed2004}
B. A. Malomed, H. E. Nistazakis, D. J. Frantzeskakis, P. G. Kevrekidis,
Static and rotating domain-wall cross patterns in Bose-Einstein condensates,
{\em Phys. Rev. A} {\bf 70} (2004) 043616.

\bibitem{Brazhnyi2005}
V. A. Brazhnyi, V. V. Konotop,
Stable and unstable vector dark solitons of coupled nonlinear Schr\"{o}dinger equations: Application to two-component Bose-Einstein condensates,
{\em Phys. Rev. E} {\bf 72} (2005) 026616.

\bibitem{Kevrekidis2005}
P. G. Kevrekidis, H. Susanto, R. Carretero-Gonz\'{a}lez, B. A. Malomed, D. J. Frantzeskakis,
Vector solitons with an embedded domain wall,
{\em Phys. Rev. E} {\bf 72} (2005) 066604.

\bibitem{Victor2005}
V\'{i}ctor M. P\'{e}rez-Garc\'{i}a, Juan Belmonte Beitia,
Symbiotic solitons in heteronuclear multicomponent Bose-Einstein condensates,
{\em Phys. Rev. A} {\bf 72} (2005) 033620.

\bibitem{Adhikari2005}
Sadhan K. Adhikari,
Bright solitons in coupled defocusing NLS equation supported by coupling: Application to Bose-Einstein condensation,
{\em Phys. Lett. A} {\bf 346} (2005) 179.

\bibitem{Xue2008}
J.-K. Xue, G.-Q. Li, A.-X. Zhang, P. Peng,
Nonlinear mode coupling and resonant excitations in two-component Bose-Einstein condensates,
{\em Phys. Rev. E} {\bf 77} (2008) 016606.


\bibitem{Ho1996}
T. L. Ho, V. B. Shenoy,
Binary Mixtures of Bose Condensates of Alkali Atoms,
{\em Phys. Rev. Lett.} {\bf 77} (1996) 3276.

\bibitem{Esry1997}
B. D. Esry, C. H. Greene, J. P. Burke, J. L. Bohn,
Hartree-Fock Theory for Double Condensates,
{\em Phys. Rev. Lett.} {\bf 78} (1997) 3594.


\bibitem{Liu2009}
X. Liu, H. Pu, B. Xiong, W. M. Liu, J. Gong,
Formation and transformation of vector solitons in two-species Bose-Einstein condensates with a tunable interaction,
{\em Phys. Rev. A} {\bf 79} (2009) 013423.

\bibitem{Zhang2009}
X.-F. Zhang, X.-H. Hu, X.-X. Liu, W. M. Liu,
Vector solitons in two-component Bose-Einstein condensates with tunable interactions and harmonic potential,
{\em Phys. Rev. A} {\bf 79} (2009) 033630.

\bibitem{Yakimenko2012}
A. I. Yakimenko, K. O. Shchebetovska, S. I. Vilchinskii, M. Weyrauch,
Stable bright solitons in two-component Bose-Einstein condensates,
{\em Phys. Rev. A} {\bf 85} (2012) 053640.

\bibitem{Wang2013}
Y.-F. Wang, B. Tian, M. Wang,
Bell-Polynomial Approach and Integrability for the Coupled Gross-Pitaevskii Equations in Bose-Einstein Condensates,
{\em Stud. Appl. Math.} {\bf 131} (2013) 119-134.


\bibitem{Bender1998}
C. M. Bender, S. Boettcher,
Real spectra in non-Hermitian Hamiltonians having $\PT$ symmetry,
{\em Phys. Rev. Lett.} {\bf 80} (1998) 5243-5246.

\bibitem{Bender2003}
C. M. Bender, D. C. Brody, H. F. Jones,
Must a Hamiltonian be Hermitian?,
{\em Am. J. Phys.} {\bf 71} (2003) 1095-1102.

\bibitem{Bender2007}
C. M. Bender,
Making sense of non-Hermitian Hamiltonians,
{\em Rep. Prog. Phys.} {\bf 70} (2007) 947-1018.


\bibitem{Nixon2012}
S. Nixon, L. Ge, J. Yang,
Stability analysis for solitons in $\PT$-symmetric optical lattices,
{\em Phys. Rev. A} {\bf 85} (2012) 023822.

\bibitem{Lumer2013}
Y. Lumer, Y. Plotnik, M. C. Rechtsman, M. Segev,
Nonlinearly Induced $\PT$ Transition in Photonic Systems,
{\em Phys. Rev. Lett.} {\bf 111}, (2013) 263901.

\bibitem{Yang2014}
J. Yang,
Symmetry breaking of solitons in one-dimensional parity-time symmetric optical potentials,
{\em Opt. Lett.} {\bf 39} (2014) 5547.


\bibitem{Ahmed2001}
Z. Ahmed,
Real and complex discrete eigenvalues in an exactly solvable one-dimensional complex $\PT$-invariant potential,
{\em Phys. Lett. A.} {\bf 282} (2001) 343-348.


\bibitem{Makris2008}
Z. H. Musslimani, K. G. Makris, R. El-Ganainy, D. N. Christodoulides,
Optical solitons in $\PT$ Periodic Potentials,
{\em Phys. Rev. Lett.} {\bf 100} (2008) 030402.

\bibitem{Yan2015}
Z. Yan, Z. Wen, C. Hang,
Spatial solitons and stability in self-focusing and defocusing Kerr nonlinear media with generalized parity-time-symmetric Scarf-II potentials,
{\em  Phys. Rev. E.} {\bf 92} (2015) 022913.

\bibitem{yan2015}
Z. Yan, Z. Wen, V. V. Konotop,
Solitons in a nonlinear Schr\"{o}dinger equation with $\PT$-symmetric potentials and inhomogeneous nonlinearity: Stability and excitation of nonlinear modes,
{\em  Phys. Rev. A.} {\bf 92} (2015) 023821.

\bibitem{Chen2018}
Y. Chen, Z. Yan, W. Liu,
Impact of near-$\PT$ symmetry on exciting solitons and interactions based on a complex Ginzburg-Landau model,
{\em  Opt. Express.} {\bf 26} (2018) 33022-33034.






\bibitem{Shi2011}
Z. Shi, X. Jiang, X. Zhu, H. Li,
Bright spatial solitons in defocusing Kerr media with $\PT$-symmetric potentials,
{\em Phys. Rev. A} {\bf 84} (2011) 053855.

\bibitem{Hu2012}
S. Hu, W. Hu,
Optical solitons in the parity-time-symmetric Bessel complex potential,
{\em J. Phys. B} {\bf 45} (2012) 225401.

\bibitem{Khare2012}
A. Khare, S. M. Al-Marzoug, H. Bahlouli,
Solitons in $\PT$-symmetric potential with competing nonlinearity,
{\em Phys. Lett. A} {\bf 376} (2012) 2880.

\bibitem{Achilleos2012}
V. Achilleos, P. G. Kevrekidis, D. J. Frantzeskakis, R. Carretero-Gonzalez,
Dark solitons and vortices in $\PT$-symmetric nonlinear media: from spontaneous symmetry breaking to nonlinear $\PT$ phase transitions,
{\em Phys. Rev. A} {\bf 86} (2012) 013808.

\bibitem{Yan2013}
Z. Y. Yan,
Complex $\PT$-symmetric nonlinear Schr\"{o}dinger equation and Burgers equation,
{\em Philos. Trans. R. Soc. Lond. A} {\bf 371} (2013) 20120059.

\bibitem{Shen2018}
Y. Shen, Z. Wen, Z. Yan, C. Hang,
Effect of $\PT$ symmetry on nonlinear waves for three-wave interaction models in the quadratic nonlinear media,
{\em Chaos} {\bf 28} (2018) 043104.

\bibitem{Xu2020}
W.-X. Xu, S.-J. Su, B. Xu, Y.-W. Guo, S.-L. Xu, Y. Zhao, Y.-H. Hu,
Two dimensional spacial soliton in atomic gases with $\PT$-symmetry potential,
{\em Opt. Express} {\bf 28} (2020) 35297.

\bibitem{Song2022}
J. Song, Z. Zhou, W. Weng, Z. Yan,
$\PT$-symmetric peakon solutions in self-focusing/defocusing power-law nonlinear media: Stability, interactions and adiabatic
excitations,
{\em Physica D} {\bf 435} (2022) 133266.

\bibitem{Zhong2023}
M. Zhong, Z. Yan, S.-F. Tian,
Stable matter-wave solitons, interactions, and excitations in the spinor $F = 1$ Bose-Einstein condensates with $\PT$-and non-$\PT$-symmetric potentials,
{\em Commun. Nonlinear Sci. Numer. Simulat.} {\bf 118} (2023) 107061.


\bibitem{Trefethen2000}
L. N. Trefethen,
Spectral methods in matlab,
SIAM, Philadelphia, 2000.

\bibitem{yang2010}
J.\ Yang,
Nonlinear Waves in Integrable and Nonintegrable Systems,
SIAM, Philadelphia, 2010.

\end{small}
\end{thebibliography}
\end{document}